\documentclass{JHEP3}
\usepackage{subfigure}
\usepackage{psfrag}
\usepackage{amsmath}
\usepackage[dvips]{epsfig}

\preprint{DAMTP-2008-31}

\title{Bayesian Selection of sign $\mu$ within mSUGRA in Global Fits Including
  WMAP5 Results}   
\author{Farhan Feroz\\
	  Cavendish Laboratory, JJ Thomson Avenue, Cambridge CB3 0HE, UK \\
        E-mail: \email{f.feroz@mrao.cam.ac.uk}}  
\author{Benjamin C Allanach\\ 
	  DAMTP, CMS, Wilberforce Road, Cambridge CB3 0WA, UK \\
        E-mail: \email{b.c.allanach@damtp.cam.ac.uk}}  
\author{Mike Hobson\\
	  Cavendish Laboratory, JJ Thomson Avenue, Cambridge CB3 0HE, UK \\
        E-mail: \email{mph@mrao.cam.ac.uk}}
\author{Shehu S AbdusSalam\\
	  DAMTP, CMS, Wilberforce Road, Cambridge CB3 0WA, UK \\
        E-mail: \email{s.s.abdussalam@damtp.cam.ac.uk}}
\author{Roberto Trotta\\
	  Astrophysics Department, Oxford University, Denys Wilkinson Building, Keble Road, Oxford OX1 3RH, UK \\
        E-mail: \email{rxt@astro.ox.ac.uk}}
\author{Arne M Weber\\
	  Max Planck Inst.\ f\"{u}r Phys., F\"{o}hringer Ring 6, D-80805 Munich, Germany \\
        E-mail: \email{arne.weber@mppmu.mpg.de}}

\abstract{We study the properties of the constrained minimal supersymmetric standard model (mSUGRA) by performing
fits to updated indirect data, including the relic density of dark matter inferred from WMAP5. In order to find
the extent to which $\mu < 0$ is disfavoured compared to $\mu > 0$, we compare the Bayesian evidence values for
these models, which we obtain straightforwardly and with good precision from the recently developed multi--modal
nested sampling (`{\sc MultiNest}') technique. We find weak to moderate evidence for the $\mu > 0$ branch of
mSUGRA over $\mu<0$ and estimate the ratio of probabilities to be $P(\mu>0)/P(\mu<0)=6-61$ depending on the
prior measure and range used. There is thus positive (but not overwhelming) evidence that $\mu>0$ in mSUGRA\@.
The {\sc MultiNest} technique also delivers probability distributions of parameters and other relevant quantities
such as superpartner masses. We explore the dependence of our results on the choice of the prior measure used. We
also use the Bayesian evidence to quantify the consistency between the mSUGRA parameter inferences coming from
the constraints that have the largest effects: $(g-2)_\mu$, $BR(b \rightarrow s \gamma)$ and cold dark matter
(DM) relic density $\Omega_{\mathrm{DM}}h^2$.}


\begin{document}

\label{firstpage}

\section{Introduction}\label{sec:intro}

The impending start of operation of the Large Hadron Collider (LHC) makes this a very exciting time for
supersymmetric (SUSY) phenomenology. Numerous groups have been pursuing a programme to fit simple SUSY models and
identify the regions in the parameter space that might be of interest with the forthcoming LHC data
\cite{baer03,Ellis:2003si,Profumo:2004at,Baltz:2004aw,Ellis:2004tc,Stark:2005mp}. The Minimal Supersymmetric Standard
Model (MSSM) with one particular choice of universal boundary conditions at the grand unification scale, called
either the Constrained Minimal Supersymmetric Standard Model (CMSSM) or mSUGRA~\cite{Arnowitt:1992aq}, has been
studied quite extensively in multi--parameter scans. mSUGRA has proved to be a popular choice for SUSY
phenomenology because of the small number of free parameters. In mSUGRA, the scalar mass $m_0$, gaugino mass
$M_{1/2}$ and tri--linear coupling $A_0$ are assumed to be universal at a gauge unification scale $M_{\rm GUT}
\sim 2 \times 10^{16}$ GeV. In addition, at the electroweak scale one selects $\tan\beta$, the ratio of Higgs
vacuum expectation values and sign$(\mu)$, where $\mu$ is the Higgs/higgsino mass parameter whose square is
computed from the potential minimisation conditions of electroweak symmetry breaking (EWSB) and the empirical
value of the mass of the $Z^0$ boson, $M_Z$. The family universality assumption is well motivated since flavour
changing neutral currents are observed to be rare. Indeed several string models (see, for example
Ref.~\cite{unistringmodels}) predict approximate MSSM universality in the soft terms. Nevertheless, mSUGRA is
just one (albeit popular) choice among a multitude of possibilities. 

Recently, Bayesian parameter estimation techniques using the Markov Chain Monte Carlo (MCMC) sampling have been
applied to the study of mSUGRA, performing a multi--dimensional Bayesian fit to indirect constraints
\cite{Allanach:2005kz, roszkowski07, deAustri:2006pe, darkSide, Allanach:2006jc, rosz:2007, Allanach:2007,
Allanach:2008iq}. Also, a study has been extended to large volume string compactified 
models~\cite{Allanach:2008tu}. A particularly important constraint comes from the cold dark matter (DM) relic
density $\Omega_{\mathrm{DM}}h^2$ determined by the Wilkinson Microwave Anisotropy Probe (WMAP). DM is assumed to
consist solely of the lightest supersymmetric particle (LSP). As pointed out in \cite{darkSide}, the accuracy of
the DM constraint results in very narrow steep regions of degenerate $\chi^2$ minima as the system is rather
under--constrained. This makes the global fit to all the relevant mSUGRA parameters potentially difficult. If the
MSSM is confirmed in the forthcoming collider data, it will hopefully be possible to break many of these
degeneracies using collider observables such as edges in kinematical distributions. However, it is expected that
one degeneracy will remain from LHC data in the form of the overall mass scale of the sparticles. We apply the
newly developed {\sc MultiNest} technique \cite{Feroz,MultiNest} to explore this highly degenerate parameter
space efficiently. With this technique, one can also calculate the `Bayesian evidence' which plays the central
role in Bayesian model selection and hence allows one to distinguish between different models.

Ref.~\cite{Heinemeyer:2008fb} performed a random scan of $10^5$ points in the parameter spaces of mSUGRA, minimal
anomaly mediated SUSY breaking (mAMSB) and minimal gauge mediated SUSY breaking (mGMSB). $b$ and electroweak
physics observables (but not the dark matter relic density) were used to assign a $\chi^2$ to each of the
points. The resulting minimum $\chi^2$ values for each scenario were then compared in order to select which
model is preferred by the data. Unfortunately, the conclusions drawn (that mAMSB is preferred by data) may have
been reversed had the dark matter relic density been included in the $\chi^2$ fit. It is also not clear how
accurate the resulting value of minimum $\chi^2$ is in each scenario, since the scans are necessarily sparse due
to the high dimensionality of the parameter space\footnote{However, this point could be easily fixed by the
authors of Ref.~\cite{Heinemeyer:2008fb} by separating the points randomly into two equally sized samples and
examining the $\chi^2$ difference of the minimum point in each.}. Recently, several studies of the mSUGRA
parameter space have used Markov Chain Monte Carlo in order to focus on the joint analysis of indirect
constraints from experiment with the $\Omega_{\mathrm{DM}}h^2$ constraint as determined by WMAP and other data.
We extend this approach by using {\sc MultiNest} to calculate the Bayesian evidence, which, when compared with
fits to different models, can be used for hypothesis testing. As an example, we consider $\mu>0$ mSUGRA versus 
$\mu<0$ mSUGRA as alternative hypotheses. In Ref.~\cite{darkSide}, the evidence ratio for these two quantities
was calculated using the method of bridge sampling~\cite{radford} in MCMCs. However, it is not clear how
accurate the estimation of the evidence ratio was, and no uncertainties were quoted. The present approach yields,
robustly small uncertainties on the ratio, for a given hypothesis and prior probability distribution. Since
Ref.~\cite{darkSide}, a tension has developed between the constraints coming from the anomalous magnetic moment
of the muon $(g-2)_\mu$, and the branching ratio of the decay of $b$ quarks into $s$ quarks $BR(b \rightarrow s
\gamma)$, which favour opposite signs of $\mu$~\cite{rosz:2007}. Ref.~\cite{rosz:2007} investigated the
constraints on continuous parameters for either sign of $\mu$ and used the Bayesian calibrated {\it p}--value
method \cite{gordon07} to get a rough estimate of the upper limit for the evidence ratio between $\mu>0$ mSUGRA
and $\mu<0$ mSUGRA of $10:1$. We also use the evidence to examine quantitatively any incompatibilities between
mSUGRA parameter inferences coming from three main constraints: $(g-2)_\mu$, $BR(b \rightarrow s \gamma)$ and 
$\Omega_{\mathrm{DM}}h^2$. Thus we determine to what extent the three measurements are compatible with each
other in an mSUGRA context. We also update the fits to WMAP5 data for the first time and include additional
$b$--physics constraints. Recent data point to an increased statistical significance in the discrepancy between
the Standard Model prediction and the experimental value of $(g-2)_\mu$, and this leads to an additional
statistical pull towards a larger contribution of $(g-2)_\mu$ coming from supersymmetry.

Our purpose in this paper is two--fold: as well as producing interesting physical insights, we also aim to gain
experience in developing and applying tools for efficient Bayesian inference, which will prove useful in the
analysis of future collider data. 

This paper is organised as follows. In Section~\ref{sec:bayesian} we motivate the case for Bayesian model 
selection. In Section~\ref{sec:analysis} we outline our theoretical setup and present our results in
Section~\ref{sec:results}. Finally, in Section~\ref{sec:summary} we list the summary and present our conclusions.
We motivate the case for the use of Bayesian evidence in quantifying consistency between different data--sets in
Appendix~\ref{app:cosistency}.

\section{Bayesian Inference}\label{sec:bayesian}

A common problem in data analysis is to use the data to make inferences about parameters of a given model. A
higher level of inference is to decide between two or more competing models. For instance, in the case of mSUGRA,
one would like to know whether there is sufficient evidence in the data to rule out the $\mu<0$ branch. Bayesian
inference provides a consistent approach to model selection as well as to the estimation of a set parameters 
$\mathbf{\Theta}$ in a model (or hypothesis) $H$ for the data $\mathbf{D}$. It can also be shown that Bayesian
inference is the unique consistent generalisation of the Boolean algebra \cite{Cox}. 

Bayes' theorem states that
\begin{equation} \Pr(\mathbf{\Theta}|\mathbf{D}, H) =
\frac{\Pr(\mathbf{D}|\,\mathbf{\Theta},H)\Pr(\mathbf{\Theta}|H)}
{\Pr(\mathbf{D}|H)},
\end{equation}
where $\Pr(\mathbf{\Theta}|\mathbf{D}, H) \equiv P(\mathbf{\Theta})$ is the posterior probability distribution
of the parameters, $\Pr(\mathbf{D}|\mathbf{\Theta}, H) \equiv \mathcal{L}(\mathbf{\Theta})$ is the likelihood,
$\Pr(\mathbf{\Theta}|H) \equiv \pi(\mathbf{\Theta})$ is the prior distribution, and $\Pr(\mathbf{D}|H) \equiv
\mathcal{Z}$ is the Bayesian evidence.

Bayesian evidence is simply the factor required to normalise the posterior over $\mathbf{\Theta}$ and is given by:
\begin{equation}
\mathcal{Z} =
\int{\mathcal{L}(\mathbf{\Theta})\pi(\mathbf{\Theta})}d^N\mathbf{\Theta},
\label{eq:3}
\end{equation}
where $N$ is the dimensionality of the parameter space. Since the Bayesian evidence does not depend on the
parameter values $\mathbf{\Theta}$, it is usually ignored in parameter estimation problems and the posterior
inferences are obtained by exploring the un--normalized posterior using standard MCMC sampling methods.

A useful feature of Bayesian parameter estimation is that one can easily obtain the posterior distribution
of any function, $f$, of the model parameters $\mathbf{\Theta}$. Since,
\begin{equation}
\Pr(f|\mathbf{D}) = \int{\Pr(f,\mathbf{\Theta} | \mathbf{D}) d \mathbf{\Theta}} = 
\int{\Pr(f|\mathbf{\Theta},\mathbf{D}) \Pr(\mathbf{\Theta} | \mathbf{D}) d \mathbf{\Theta}}
= \int{\delta(f(\mathbf{\Theta})-f) \Pr(\mathbf{\Theta}|\mathbf{D}) d \mathbf{\Theta}}
\label{eq:derived_post}
\end{equation} 
where $\delta(x)$ is the delta function. Thus one simply needs to compute $f(\mathbf{\Theta})$ for every Monte
Carlo sample and the resulting sample will be drawn from $\Pr(f|\mathbf{D})$. We make use of this feature in
Section~\ref{sec:const} where we present the posterior probability distributions of various observables used in the
analysis of mSUGRA model.

In order to select between two models $H_{0}$ and $H_{1}$ one needs to compare their respective posterior
probabilities given the observed data set $\mathbf{D}$, as follows:
\begin{equation}
\frac{\Pr(H_{1}|\mathbf{D})}{\Pr(H_{0}|\mathbf{D})}
=\frac{\Pr(\mathbf{D}|H_{1})\Pr(H_{1})}{\Pr(\mathbf{D}|
H_{0})\Pr(H_{0})}
=\frac{\mathcal{Z}_1}{\mathcal{Z}_0}\frac{\Pr(H_{1})}{\Pr(H_{0})},
\label{eq:3.1}
\end{equation}
where $\Pr(H_{1})/\Pr(H_{0})$ is the prior probability ratio for the two models, which can often be set to unity
but occasionally requires further consideration. It can be seen from Eq.~\ref{eq:3.1} that the Bayesian evidence
takes the center stage in Bayesian model selection. As the average of likelihood over the prior, the Bayesian
evidence is higher for a model if more of its parameter space is likely and smaller for a model with highly
peaked likelihood but has many regions in the parameter space with low likelihood values. Hence, Bayesian model
selection automatically implements Occam's razor: a simpler theory which agrees well enough with the empirical
evidence is preferred. A more complicated theory will only have a higher evidence if it is significantly better
at explaining the data than a simpler theory.

Unfortunately, evaluation of the multidimensional integral \eqref{eq:3} is a challenging numerical task. Standard
techniques like thermodynamic integration \cite{O'Ruanaidh} are extremely computationally expensive which makes
evidence evaluation typically at least an order of magnitude more costly than parameter estimation. Some fast
approximate methods have been used for evidence evaluation, such as treating the posterior as a multivariate
Gaussian centred at its peak (see e.g. Ref.~\cite{Hobson}), but this approximation is clearly a poor one for
multi--modal posteriors (except perhaps if one performs a separate Gaussian approximation at each mode). The
Savage--Dickey density ratio has also been proposed \cite{savageDickey} as an exact, and potentially faster,
means of evaluating evidences, but is restricted to the special case of nested hypotheses and a separable prior
on the model parameters. Bridge sampling~\cite{radford} allows the evaluation of the ratio of Bayesian evidence
of two models and is implemented in the `bank sampling' method of Ref.~\cite{bank} but it is not yet clear how
accurately bank sampling can calculate these evidence ratios. Various alternative information criteria for model
selection are discussed by \cite{Liddle}, but the evidence remains the preferred method.

The nested sampling approach, introduced by Skilling \cite{Skilling}, is a Monte Carlo method targeted at the 
efficient calculation of the evidence, but also produces posterior inferences as a by--product. Feroz \& Hobson
\cite{Feroz,MultiNest} built on this nested sampling framework and have recently introduced the {\sc MultiNest}
algorithm which is efficient in sampling from multi--modal posteriors exhibiting curving degeneracies, producing
posterior samples and calculating the evidence value and its uncertainty. This technique has greatly reduced the
computational cost of model selection and the exploration of highly degenerate multi--modal posterior
distributions. We employ this technique in this paper.

The natural logarithm of the ratio of posterior model probabilities provides a useful guide to what constitutes a
significant difference between two models:

\begin{equation}
\log \Delta E = \log \left[ \frac{\Pr(H_{1}|\mathbf{D})}{\Pr(H_{0}|\mathbf{D})}\right]
=\log \left[ \frac{\mathcal{Z}_1}{\mathcal{Z}_0}\frac{\Pr(H_{1})}{\Pr(H_{0})}\right].
\label{eq:Jeffreys}
\end{equation}
We summarise convention we use in this paper in Table~\ref{tab:Jeffreys}.

\TABULAR{|l|l|l|l|}{\hline
$|\log \Delta E|$ & Odds & Probability & Remark \\ \hline
$<1.0$ & $\lesssim 3:1$ & $<0.750$ & Inconclusive \\
$1.0$ & $\sim 3:1$ & $0.750$ & Weak Evidence \\
$2.5$ & $\sim 12:1$ & $0.923$ & Moderate Evidence \\
$5.0$ & $\sim 150:1$ & $0.993$ & Strong Evidence \\ \hline
}{The scale we use for the interpretation of model probabilities. Here the $\log$
represents the natural logarithm. \label{tab:Jeffreys}}

While for parameter estimation, the priors become irrelevant once the data are powerful enough, for model
selection the dependence on priors always remains (although with more informative data the degree of dependence
on the priors is expected to decrease, see e.g. Ref.~\cite{trotta08}); indeed this explicit dependence on priors
is one of the most attractive features of Bayesian model selection. Priors should ideally represent one's state
of knowledge before obtaining the data. Rather than seeking a unique `right' prior, one should check the
robustness of conclusions under reasonable variation of the priors. Such a sensitivity analysis is required to
ensure that the resulting model comparison is not overly dependent on a particular choice of prior and the
associated metric in parameter space, which controls the value of the integral involved in the computation of the
Bayesian evidence (for some relevant cautionary notes on the subject see Ref.~\cite{Cousins:2008gf}).

One of the most important applications of model selection is to decide whether the introduction of new parameters
is necessary. Frequentist approaches revolve around the significance test and goodness--of--fit statistics, where
one accepts the additional parameter based on the improvement in $\Delta\chi^2$ by some chosen threshold. It has
been shown that such tests can be misleading (see e.g. Ref.~\cite{savageDickey,gordon07}), not least because they
depend only on the values of $\chi^2$ at the best--fit point, rather than over the entire allowed range of  the
parameters. 

Another application of Bayesian model selection is in quantifying the consistency between two or more data sets
or constraints \cite{Hobson,consistency}. Different experimental observables may ``pull'' the model parameters
in different directions and consequently favour different regions of the parameter space. Any obvious conflicts
between the observables are likely to be noticed by the ``chi by eye'' method employed to date but it is
imperative for forthcoming high--quality constraints to have a method that can quantify these discrepancies. The
simplest scenario for analysing different constraints on a particular model is to assume that all constraints
provide information on the same set of parameter values. We represent this hypothesis by $H_1$. This is the
assumption which underlies the joint analysis of the constraints. However, if we are interested in accuracy as
well as precision then any systematic differences between constraints should also be taken into account. In the
most extreme case, which we represent by $H_0$, the constraints would be in conflict to such an extent that
each constraint requires its own set of parameter values, since they are in different regions of parameter space.
Bayesian evidence provides a very easy method of distinguishing between scenarios, $H_0$ and $H_1$. To see this,
we again make use of Eq.~\ref{eq:3.1}. If we have no reason to favour either of $H_0$ or $H_1$ over the
other, then we can distinguish between these two scenarios using the following ratio,
\begin{equation}
R=\frac{\Pr(\mathbf{D}|H_{1})}{\Pr(\mathbf{D}|H_{0})}=\frac{\Pr(\mathbf{D}|H_{1})}{\prod_{i}^{}\Pr(D_i|H_{0})}. 
\label{eq:consistency}
\end{equation}
Here the numerator represents the joint analysis of all the constraints $\mathbf{D}=\{D_1,D_2,\ldots,D_n\}$ while
in the denominator the individual constraints $D_1,D_2,\ldots,D_n$ are assumed to be independent and are each
fit individually to mSUGRA, with a different set of mSUGRA parameters for each $D_i$. The interpretation of the
$\log R$ value can be made in a similar manner to model selection, as discussed in the preceding paragraph. A
positive value of $\log R$ gives the evidence in favour of the hypothesis $H_1$ that all the constraints are
consistent with each other while a negative value would point towards tension between constraints, which prefer
different regions of mSUGRA parameter space. We follow this recipe to carry out consistency checks for the 
mSUGRA model between $(g-2)_\mu$, $BR(b \rightarrow s \gamma)$ and $\Omega_{\mathrm{DM}}h^2$ as determined by
WMAP and other cosmological measurements. The $H_1$ hypothesis thus states that mSUGRA jointly fits these three
observables, whereas $H_0$ states that they all prefer different regions of parameter space and so we require an
`(mSUGRA)$^3$' model to fit them. Given the fact that Bayesian evidence naturally embodies a quantification of
Occam's razor, the resulting complexity in the model coming from the additional 2 sets of mSUGRA parameters must
be matched by a better fit to data for $H_0$ to be preferred.

\section{The Analysis}\label{sec:analysis}

Our parameter space $\mathbf{\Theta}$ contains 8 parameters, 4 of them being the mSUGRA parameters; $m_0$,
$M_{1/2}$, $A_0$, $\tan\beta$ and the rest taken from the Standard Model (SM): the QED coupling constant in the
${\overline MS}$ scheme $\alpha^{\overline{MS}}(M_Z)$, the strong coupling constant
$\alpha_s^{\overline{MS}}(M_Z)$, the running mass of the bottom quark $m_b(m_b)^{\overline{MS}}$ and the pole
top mass $m_t$. We refer to these SM parameters as nuisance parameters. Experimental errors on the mass $M_Z$
of the $Z^0$ boson and the muon decay constant $G_\mu$ are so small that we fix these parameters to their
central values of $91.1876$ GeV and $1.16637 \times 10^{-5}$ GeV$^{-2}$ respectively.

For all the models analysed in this paper, we used 4,000 live points (see Refs.~\cite{Feroz,MultiNest}) with the
{\sc MultiNest} technique. This corresponds to around 400,000 likelihood evaluations taking approximately $20$
hours on 4 3.0 GHz Intel Woodcrest processors.

\subsection{The Choice of Prior Probability Distribution}\label{sec:analysis:prior} 
In all cases, we assume the prior is separable, such that
\begin{equation}
\pi(\mathbf{\Theta})=\pi(\theta_1)\pi(\theta_2) \ldots \pi(\theta_8),
\label{eq:sep_prior}
\end{equation}
where $\pi(\theta_i)$ represents the prior probability of parameter $\theta_i$. We consider two initial ranges
for the mSUGRA parameters which are listed in Table~\ref{tab:range}. 
\TABULAR{|l|c|c|}{\hline
mSUGRA parameters & 2 TeV range & 4 TeV range \\ \hline
$m_0$ & 60 GeV to 2 TeV & 60 GeV to 4 TeV \\
$M_{1/2}$ & 60 GeV to 2 TeV & 60 GeV to 4 TeV \\
$A_0$ & --4 TeV to 4 TeV & --7 TeV to 7 TeV \\
$\tan\beta$ & 2 to 62 & 2 to 62 \\ \hline
}{mSUGRA uniform prior parameter ranges \label{tab:range}}
The ``2 TeV'' range is motivated by a general ``naturalness'' argument that SUSY mass parameters should lie
within $\mathcal{O}(1$ TeV), since otherwise a fine-tuning in the electroweak symmetry breaking sector results.
Deciding which region of parameter space is natural is obviously subjective. For this reason, we include the
``4 TeV'' range results to check the dependence on prior ranges. We consider the branches $\mu < 0$ and $\mu >
0$ separately.

\TABULAR{|l|cc|c|}{\hline
SM parameters & Mean value & Uncertainty & Reference \\
& $\mu$ & $\sigma$ (exp) &  \\ \hline
$1/\alpha^{\overline{MS}}$ & $127.918$ & $0.018$ & \cite{pdg06} \\
$\alpha_s^{\overline{MS}}(M_Z)$ & $0.1176$ & $0.002$ & \cite{pdg06} \\
$m_b(m_b)^{\overline{MS}}$ & $4.20$ GeV & $0.07$ GeV & \cite{pdg06} \\
$m_t$ & $170.9$ GeV & $1.8$ GeV & \cite{cdf+dzero-mtop-07}\\ \hline
}{Constraints on the Standard Model (nuisance) parameters \label{tab:SM}}
We impose flat priors on all 4 mSUGRA parameters (i.e. $m_0, M_{1/2}, A_0$ and $\tan\beta$) for the ``2 TeV'' and
``4 TeV'' ranges and both signs of $\mu$. Current constraints on SM (nuisance) parameters are listed in 
Table~\ref{tab:SM}\footnote{We note that the experimental constraint on $m_t$ is changing quite rapidly as  new
results are issued from the Tevatron experiments. The latest combined constraint (released after this paper was
first written) is  $m_t=172.4 \pm 1.2$ GeV~\cite{newMt}. Any fit differences caused in the movement of the
central value will be smeared out by its uncertainty, but we shall mention at the relevant point below where the
new value could change the fits.}.
With the means and $1\sigma$ uncertainties from Table~\ref{tab:SM}, we impose Gaussian priors on
SM (nuisance) parameters truncated at $4\sigma$ from their central values. We also perform the analysis for 
flat priors in $\log m_0$ and $\log M_{1/2}$ for both ranges and both signs of $\mu$. Since,
\begin{equation}
  \int d \log m_0\ d \log M_{1/2}\ p(m_0, M_{1/2} | {\bf D}) =
  \int {d m_0}\ d M_{1/2}\ \frac{p(m_0, M_{1/2} | {\bf D})}{m_0 M_{1/2}}
\end{equation}
it is clear that the logarithmic prior measures have a factor $1/(m_0 M_{1/2})$ compared to the linear prior
measure and so it could potentially favour lighter sparticles. If the data constrains the model strongly enough,
lighter sparticles would only be favoured negligibly. Our main motive in seeing the variation of the fit to the
variation in prior measure is to check the dependence of our results on the choice of the prior. For robust
fits, which occur when there is enough precisely constraining data, the posterior probability density should
only have a small dependence upon the precise form of the prior measure.

\subsection{The Likelihood}\label{sec:analysis:likelihood}

\TABULAR{|l | c c | c|}{\hline
Observable &  Mean value & Uncertainty & Reference \\\hline
${\delta a_{\mu}} \times 10^{-10}$ & $29.5$ & $8.8$ & \cite{gm2SM}\\
$M_W$ & $80.398$ GeV & $27$ MeV & \cite{mw,Heinemeyer:2006px} \\
$\sin^2 \theta_w^l$ & $0.23149$ & $0.000173$ &  \cite{sinth,mw}  \\
$BR(b \rightarrow s \gamma)\times 10^{4}$ & $3.55$ & $0.72$ & \cite{superiso,gamb}\\
$\Delta_{o-}$ & $0.0375$ & $0.0289$ & \cite{superiso,gamb} \\
$R_{BR(B_u \rightarrow \tau \nu)}$ & $1.259$ & $0.378$ & \cite{btaunu} \\
$R_{\Delta_{m_s}}$ & $0.85$ & $0.12$ & \cite{btaunu,delms_sm} \\ \hline
}{Summary of the Gaussian distributed observables used in the analysis. For each quantity we use a likelihood
function with central mean $\mu$ and standard deviation $s=\sqrt{\sigma^2+\tau^2}$ where $\sigma$ is the
experimental uncertainty and $\tau$ is the theoretical uncertainty. $\Delta_{o-}$ represents the isospin
asymmetry of $B \rightarrow K^* \gamma$. $R_{BR(B_u \rightarrow \tau \nu)}$ represents the ratio of the
experimental and SM predictions of the branching ratio of $B_u$ mesons decaying into a tau and a tau neutrino.
$R_{\Delta_{m_s}}$ is the ratio of the experimental and the SM neutral $B_s$ meson mixing amplitudes. The
non-Gaussian likelihoods for the LEP constraint on Higgs mass, $BR(B_s\rightarrow \mu^+\mu^-)$ and 
$\Omega_{\mathrm{DM}} h^2$ are described later.
\label{tab:observables}}

Our calculation of the likelihood closely follows Ref.~\cite{Allanach:2007}, with updated data and additional
variables included, and is summarised in Table~\ref{tab:observables} and
discussed further below. 
We assume that
the measurements $D_i$ of observables (the `data') used in our likelihood calculation are independent and have
Gaussian errors\footnote{The experimental constraints the LEP constraint on Higgs mass, $BR(B_s\rightarrow
\mu^+\mu^-)$ and $\Omega_{\mathrm{DM}} h^2$ likelihood, each described later, are not Gaussian.}, so that the
likelihood distribution for a given model ($H$) is
\begin{equation}
\mathcal{L}(\mathbf{\Theta}) \equiv \Pr(\mathbf{D}|\mathbf{\Theta}, H) = \prod_{i}^{} \Pr(D_i|\mathbf{\Theta}, H),
\label{eq:like1}
\end{equation}
where
\begin{equation}
\Pr(D_i|\mathbf{\Theta}, H)=\frac{1}{\sqrt{2\pi \sigma_i^2}}\exp[-\chi^2/2]
\label{eq:likelihood}
\end{equation}
and
\begin{equation}
\chi^2=\frac{(c_i-p_i)^2}{\sigma_i^2}.
\label{eq:like2}
\end{equation}
$p_i$ is the ``predicted'' value of the observable $i$ given the knowledge of the model $H$ and $\sigma$ is the
standard error of the measurement.

In order to calculate predictions $p_i$ for observables from the input parameters $\mathbf{\Theta}$, {\tt
SOFTSUSY2.0.17}~\cite{softsusy} is first employed to calculate the MSSM spectrum. Bounds upon the sparticle
spectrum have been updated and are based upon the bounds collected in Ref.~\cite{deAustri:2006pe}. Any spectrum
violating a 95$\%$ limit from negative sparticle searches is assigned a zero likelihood density. Also, we set a
zero likelihood for any inconsistent point, e.g.\ one which does not break electroweak symmetry correctly, has a
charged LSP, or has tachyonic sparticles. For points that are not ruled out, we then link the mSUGRA spectrum via
the SUSY Les Houches Accord~\cite{slha} (SLHA) to various other computer codes  that calculate various
observables. For instance,  {\tt micrOMEGAs1.3.6} \cite{micromegas}, calculates  $\Omega_{\mathrm{DM}} h^2$, the
branching ratio $BR(B_s \rightarrow \mu^+ \mu^-)$ and the anomalous magnetic moment of the muon $(g-2)_\mu$.

The anomalous magnetic moment of the muon $a_\mu\equiv(g-2)_\mu/2$ was  measured to be  
$a^\mathrm{\mathrm{exp}}_\mu=(11659208.0\pm6.3)\times 10^{-10}$ \cite{gm2exp}. Its experimental value is in
conflict with the SM predicted value $a_\mu^{\mathrm{SM}}=(11659178.5\pm6.1)\times10^{-10}$ from~\cite{gm2SM},
which includes the latest QED~\cite{gm2QED}, electroweak~\cite{gm2EW}, and hadronic~\cite{gm2SM} contributions to
$a^{\mathrm{SM}}_\mu$. This SM prediction does not however account for $\tau$ data which is known to lead to
significantly different results for $a_\mu$, implying underlying theoretical difficulties which have not been
resolved so far. Restricting to $e^+e^-$ data, hence using the numbers given above, we find
\begin{equation}
 \delta \frac{(g-2)_\mu}{2}\equiv \delta a_\mu 
 \equiv a_\mu^{\mathrm{exp}}-a_\mu^{\mathrm{SM}}=(29.5 \pm 8.8)\times
 10^{-10}.
\label{g-2}
\end{equation}
This excess may be explained by a supersymmetric contribution, the sign of which is identical in mSUGRA to the
sign of the super potential $\mu$ parameter~\cite{susycont}. After obtaining the one-loop MSSM value of
$(g-2)_\mu$ from {\tt micrOMEGAs v1.3.6}, we add the dominant 2-loop corrections detailed in
Refs.~\cite{2loop,private}. 

The $W$ boson pole mass $M_W$ and the effective leptonic mixing angle $\sin^2 \theta^l_w$ are also used in the
likelihood. We take the measurements to be~\cite{mw,sinth} \begin{equation} M_W = 80.398\pm0.027\mbox{~GeV},
\qquad \sin^2 \theta_w^l = 0.23149 \pm 0.000173, \label{ewobs} \end{equation} where experimental errors and
theoretical uncertainties due to missing higher--order corrections in SM~\cite{mwsmbest} and
MSSM~\cite{Heinemeyer:2006px,drMSSMal2B} have been added in quadrature. The most up to date MSSM predictions for
$M_W$ and $\sin^2 \theta_w^l$~\cite{Heinemeyer:2006px} are finally used to compute the corresponding 
likelihoods. 

A parameterisation of the LEP2 Higgs search likelihood for various Standard Model Higgs masses is utilised, since
the lightest Higgs $h$ of mSUGRA is very SM-like once the direct search constraints are taken into account. It is
smeared with a 2 GeV assumed theoretical uncertainty in the {\tt SOFTSUSY2.0.17} prediction of $m_h$ as described
in \cite{Allanach:2006jc}. 

The experimental value of the rare bottom quark branching ratio to a strange quark and a photon $BR(b
\rightarrow s \gamma)$ is constrained to be~\cite{hfg} 
\begin{equation}
BR(b \rightarrow s \gamma)= (3.55\pm0.26) \times 10^{-4}. 
\label{bsg}
\end{equation}
The SM prediction has recently moved down quite substantially from $(3.60\pm0.30) \times 10^{-4}$ to
$(3.15\pm0.23) \times 10^{-4}$ \cite{ms06-bsg,mm-prl06}. This shift was caused by including most of the
next-to-next-to-leading order (NNLO) perturbative QCD contributions as well as the leading non-perturbative and
electroweak effects. We use the publicly available code {\tt SuperIso2.0} \cite{superiso} (linked via the SLHA to
the mSUGRA spectrum predicted) which computes $BR(b \rightarrow s \gamma)$ in the MSSM with Minimal Flavor
Violation. We note that mSUGRA is of such a minimal flavor violating form, and so the assumptions present in {\tt
SuperIso2.0} are the appropriate ones. The computation takes into account one-loop SUSY contributions, as well as
$\tan \beta$-enhanced two-loop contributions in the effective lagrangian approach. The recent partial NNLO SM QCD
corrections are also included by the program. Ref. \cite{gamb} derives a 95\% interval for the bounds including
the experimental and theory SM/MSSM errors to be
\begin{equation}
2.07 \times 10^{-4} < BR(b \rightarrow s \gamma) < 4.84 \times 10^{-4}. 
\label{bsg_bounds}
\end{equation}
For the constraint on $BR(b \rightarrow s \gamma)$, we use the mean value of $3.55\times 10^{-4}$ and derive the
1--$\sigma$ uncertainty from the above given bound to be equal to $0.72\times 10^{-4}$. We note that this is
twice as large as the uncertainty used in another recent global fit~\cite{rosz:2007}, where an enhancement in
the posterior density of the large $\tan \beta$ region was observed to result from the new constraint. 

The new upper 95$\%$ C.L. bound on $BR(B_s \rightarrow \mu^+ \mu^-)$ coming from the CDF collaboration is
5.8$\times 10^{-8}$. We are in possession~\cite{private2} of the empirical $\chi^2$ penalty for this observable
as a function of the predicted value of $BR(B_s \rightarrow \mu^+ \mu^-)$ from old CDF data when the 95$\%$ C.L.
upper bound was 0.98$\times 10^{-8}$. Here, we assume that the shape of the likelihood penalty coming from data
is the same as presented in Ref.~\cite{darkSide}, but that only the normalisation of the branching ratio shifts
by the ratio of the 95$\%$ C.L. upper bounds: $0.58/0.98$. 

For the $\Delta_{o-}$, isospin asymmetry of $B \rightarrow K^* \gamma$, the 95\% confidence level for the
experimental results from the combined BABAR and Belle data combined with the theoretical errors is \cite{gamb}:
\begin{equation}
-0.018 \times 10^{-4} < \Delta_{o-} < 0.093 \times 10^{-4},
\label{iso_bounds}
\end{equation}
with the central value of $0.0375$. We derive the 1--$\sigma$ uncertainty from the above  given bound to be equal
to $0.0289$. We use the publicly available code {\tt SuperIso2.0}  \cite{superiso} to calculate $\Delta_{o-}$. We
neglect experimental correlations between the measurements of $\Delta_{o-}$ and $BR(b \rightarrow s \gamma)$. In
practice, the $\Delta_{o-}$ constraint makes a much smaller difference than $BR(b \rightarrow s\gamma)$ to our
fits, and so we expect the inclusion of a correlation to also have a small effect. The parametric correlations
caused by variations of $\alpha_s(M_Z)$ and $m_b(m_b)$ are included by our analysis, since they are varied as
input parameters. 

The average experimental value of $BR(B_u \rightarrow \tau \nu)$ from HFAG \cite{btaunu} (under purely leptonic
modes) is:
\begin{equation}
BR^{\mathrm{exp}}(B_u \rightarrow \tau \nu) = (141 \pm 43) \times 10^{-6}.
\label{btaunu_exp}
\end{equation}
The SM prediction is rather uncertain because of two incompatible empirically derived values of $|V_{ub}|$: one
being $(3.68 \pm 0.14) \times 10^{-3}$. The other comes from inclusive semi-leptonic decays and is $(4.49 \pm
0.33) \times 10^{-3}$. These lead to $BR(B_u \rightarrow \tau \nu)$ values of $(0.85 \pm 0.13) \times 10^{-4}$
and $(1.39 \pm 0.44) \times 10^{-4}$ respectively. We statistically average these two by averaging the central
values, and then adding the errors in quadrature and dividing by $\sqrt{2}$. This gives:
\begin{equation}
BR^{\mathrm{SM}}(B_u \rightarrow \tau \nu) = (112 \pm 25) \times 10^{-6}.
\label{btaunu_pred}
\end{equation}
Taking the ratio of the experimental and SM values of $BR(B_u \rightarrow \tau \nu)$ gives:
\begin{equation}
R_{BR(B_u \rightarrow \tau \nu)} = 1.259 \pm 0.378.
\label{btaunu_ratio}
\end{equation}
For the MSSM prediction, we use the formulae in Ref.~\cite{Isidori:2006pk}, which include the large $\tan \beta$
limit of one-loop corrections coming from loops involving a charged Higgs.

The experimental and SM-predicted values of the neutral $B_s$ meson mixing amplitude are 
\cite{btaunu,delms_sm}:
\begin{equation}
\Delta^{\mathrm{exp}}m_s = 17.77 \pm 0.12 \mbox{~ps}^{-1}, \Delta^{\mathrm{SM}}m_s = 20.9 \pm 2.6 \mbox{~ps}^{-1}.
\label{eq:delms_exp_sm}
\end{equation}
Taking the ratio of these two values, we get:
\begin{equation}
R_{\Delta m_s} = 0.85 \pm 0.12.
\label{eq:delms}
\end{equation}
We use the formulae of Ref.~\cite{Buras:2002vd} for the MSSM prediction of $R_{\Delta m_S}$, calculating it in
the large $\tan \beta$ approximation. The dominant correction comes from one-loop diagrams involving a neutral
Higgs boson. 

\FIGURE{
\scalebox{1.0}{
\begin{picture}(0,0)%
\includegraphics{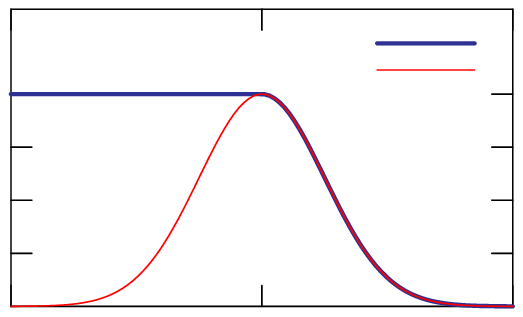}%
\end{picture}%
\setlength{\unitlength}{0.0200bp}%
\begin{picture}(10800,6480)(0,0)%
\put(2475,2414){\makebox(0,0)[r]{\strut{} 0.25}}%
\put(2475,3179){\makebox(0,0)[r]{\strut{} 0.5}}%
\put(2475,3943){\makebox(0,0)[r]{\strut{} 0.75}}%
\put(2475,4707){\makebox(0,0)[r]{\strut{} 1}}%
\put(2750,1100){\makebox(0,0){\strut{} 0.0343}}%
\put(6363,1100){\makebox(0,0){\strut{} 0.1143}}%
\put(9975,1100){\makebox(0,0){\strut{} 0.1943}}%
\put(550,3790){\rotatebox{90}{\makebox(0,0){\strut{}$L/L_{max}$}}}%
\put(6362,275){\makebox(0,0){\strut{}$\Omega_{DM} h^2$}}%
\put(7750,5438){\makebox(0,0)[r]{\strut{}constraint}}%
\put(7750,5054){\makebox(0,0)[r]{\strut{}pure WMAP5}}%
\end{picture}%
 
}
\caption{Depiction of our likelihood constraint on the predicted value of $\Omega_{\mathrm{DM}}h^2$ due to
lightest neutralinos, compared to a simple Gaussian with WMAP5 central value and a 1$\sigma$ uncertainty of
0.02.}
\label{fig:omega}}

The WMAP 5--year data combined with the distance measurements from the Type Ia supernovae (SN) and the Baryon
Acoustic Oscillations (BAO) in the distribution of galaxies gives the $\Lambda$-cold dark matter fitted value of
the dark matter relic density \cite{wmap}:
\begin{equation}
\Omega \equiv \Omega_{\mathrm{DM}} h^2 = 0.1143 \pm 0.0034.
\label{omega}
\end{equation}

In the present paper, we assume that the dark matter consists of neutralino, the LSP. Recently, it has been shown
that the LSP relic density is highly sensitive to the pre-Big Bang Nucleosynthesis (BBN) rate and even a modest
modification can greatly enhance the calculated relic density with no contradiction with the cosmological
observations \cite{BBN_sensitivity}. It is also possible that a non-neutralino component of dark matter is
concurrently present and indeed the inclusion of neutrino mass\-es via right-handed neutrinos can change the
relic density prediction somewhat \cite{barger08}. We therefore penalise only for the predicted
$\Omega_{\mathrm{DM}}h^2$ being greater than the WMAP5 + BAO + SN central value. We define $x$ to be the
predicted value of $\Omega_{\mathrm{DM}}h^2$, $c=0.1143$ to be  the central $\Omega_{\mathrm{DM}}h^2$ value from
WMAP5 + BAO + SN observations and $s$ to be the error on the predicted $\Omega_{\mathrm{DM}}h^2$ value which
includes theoretical as well as experimental components. We take $s=0.02$ in order to incorporate an estimate of
higher order uncertainties in its prediction~\cite{Baro:2007em} and we define the likelihood as:
\begin{equation}
\mathcal{L}(x \equiv \Omega_{\mathrm{DM}}h^2) = 
\begin{cases}
\frac{1}{c+\sqrt{\pi s^2/2}}, & 
\mbox{if $x < c$} \\
\frac{1}{c+\sqrt{\pi s^2/2}}\exp\left[-\frac{(x-c)^2}{2s^2}\right], & 
\mbox{if $x \geq c$}. \\
\end{cases}
\label{omega_like}
\end{equation}
A diagram of the resulting likelihood penalty is displayed in Fig.~\ref{fig:omega}. This differs slightly from
the formulation suggested previously by one of the authors, for $\mathcal{L}(x \equiv \Omega_{\mathrm{DM}}h^2)$
for the case when a non-neutralino component of dark matter is concurrently present, which drops more quickly
than our flat likelihood up until the peak of WMAP Gaussian likelihood distribution.

\section{Results}\label{sec:results}

In this section, we first show our main results on the quantification of the preference of the fits for $\mu>0$.
Next, we show some highlights of updated parameter constraints coming from the fit, finishing with a study on
the level of compatibility of various observables.
 
\subsection{Model Comparison}\label{sec:results:model}

We summarise our main results in Table~\ref{tab:prob_odds} in which we list the posterior model probability odds,
$P_+/P_-$ for mSUGRA models with $\mu > 0$ and $\mu < 0$, for the two prior ranges used with flat and logarithmic
prior measures as discussed in Section~\ref{sec:analysis}. The calculation of the ratio of posterior model
probabilities requires the prior probability ratio for the two signs of $\mu$ (see Section~\ref{sec:bayesian}),
which we have set to unity. One could easily calculate the ratio $P_+/P_-$ for a different prior probability
ratio $r$, by multiplying $P_+/P_-$ in Table~\ref{tab:prob_odds} with $r$. From the probability odds listed in
Table~\ref{tab:prob_odds}, although there is a positive evidence in favour of mSUGRA model with $\mu > 0$, the
extent of the preference depends quite strongly on the priors used and the evidence ranges from being relatively
strong in the case of logarithmic prior with ``$2$ TeV'' range to weak for flat priors with ``$4$ TeV'' range.
This dependence on the prior is a clear sign that the data are not yet of sufficiently high quality to be able to
distinguish between these models unambiguously. Hopefully, the forthcoming high-quality data from LHC would be
able to cast more light on it.

\TABULAR{|c|cc|cc|}{\hline
Prior & \multicolumn{2}{c|}{``2 TeV''} & \multicolumn{2}{c|}{``4 TeV''} \\
& flat   & log    & flat   & log    \\ \hline
$\log \Delta E$ (our determination) & $2.7 \pm 0.1$ & $4.1 \pm 0.1$ & $1.8 \pm 0.1$ & $3.2 \pm 0.1$ \\ 
$P_+/P_-$ (our determination) & $15.6 \pm 1.1$ & $61.6 \pm 4.3$ & $5.9 \pm 0.4$ & $24.0 \pm 1.7$ \\ \hline
$\log \Delta E$ (from Ref.~\protect\cite{darkSide}) & $2.1$ & $-$ & $1.8$ & $2.7$ \\ 
$P_+/P_-$ (from Ref.~\protect\cite{darkSide}) & $8.3$ & $-$ & $6.2$ & $14.3$
\\ \hline 
}{The posterior probability ratios for mSUGRA model with different signs of $\mu$. Here we have assumed the prior
probabilities of the different signs of $\mu$ to be same. The uncertainties on $\log \Delta E$ for mSUGRA model
with different signs of $\mu$ are the same for different priors, since with the {\sc MultiNest} technique, the
uncertainty on the evidence value is set by the number of live points and the stopping criteria (see
Refs.~\cite{Feroz,MultiNest}) which were the same for different priors used in this study. The second row shows,
for comparison, a previous determination with earlier data using the much less precise {\em bridge sampling}
method. Some aspects of this fit were somewhat different to the present work's approach and are discussed in the
text. \label{tab:prob_odds}}

We also show in Table~\ref{tab:prob_odds} for comparison, the probability ratio $P_+/P_-$ determined in an earlier
MCMC fit using different data~\cite{darkSide}. We can see that our determination of the probability ratio favours
$\mu>0$ more strongly than Ref.~\cite{darkSide}. The main factors affecting this are that Ref.~\cite{darkSide}
had an anomalous magnetic moment of the muon less in conflict with experiment: $\delta a_\mu = (22 \pm 10) \times
10^{-10}$ as opposed to Eq.~\ref{g-2} in the present analysis, which also includes the additional
$b$-observables: $\Delta m_s$, $BR(B_u \rightarrow \tau \nu)$ and $\Delta_{0-}$. Some other details of the fit
were also different in Ref.~\cite{darkSide}: for instance $M_{1/2}<2$ TeV for all fits, and the range of $A_0$
was different. These ranges will affect the evidence obtained, at least to some degree. Unfortunately,
Ref.~\cite{darkSide} neglects to present statistical errors in the determination of the ratios of evidence
values, a situation which is rectified in Table~\ref{tab:prob_odds}. It is clear from Table~\ref{tab:prob_odds} 
that the uncertainty in the result of the model comparison is presently dominated by the prior choice, rather
than by the small statistical uncertainty in the determination of the evidence ratio with {\sc MultiNest}. It
can however be concluded that present data favour the $\mu>0$ branch of mSUGRA with a Bayesian evidence going
from weak to moderate, depending on the choice of prior.

To quantify the extent to which these results depend on $(g-2)_\mu$ constraint, we calculate the Bayesian
evidence ratio, for mSUGRA models with $\mu > 0$ and $\mu < 0$, for the flat ``4 TeV'' range priors with all the
observables discussed in Section~\ref{sec:analysis:likelihood} apart from $(g-2)_\mu$. We find $\log \Delta E = -0.5
\pm 0.1$ translating into posterior probability odds $P_+/P_- = 0.6 \pm 1.1$. This shows that in the absence of 
$(g-2)_\mu$ constraint, both mSUGRA models with $\mu > 0$ and $\mu < 0$ are equally favoured by the data.
Inclusion of $(g-2)_\mu$ constraint causes a shift of 2.3 $\log$ units in favour of $\mu > 0$ for the linear ``4
TeV'' range prior measure and hence it can be concluded that $(g-2)_\mu$ does indeed dominate our model
selection results in favour of $\mu > 0$.

\subsection{Updated Parameter Constraints} \label{sec:const}
\FIGURE{
\psfrag{m12}{$m_{1/2}$ (TeV)}
\psfrag{m0}{$m_0$ (TeV)}
\psfrag{A0}{$A_0$ (TeV)}
\psfrag{tanb}{$\tan \beta$}
\includegraphics[width=0.4\columnwidth, height=0.4\columnwidth]{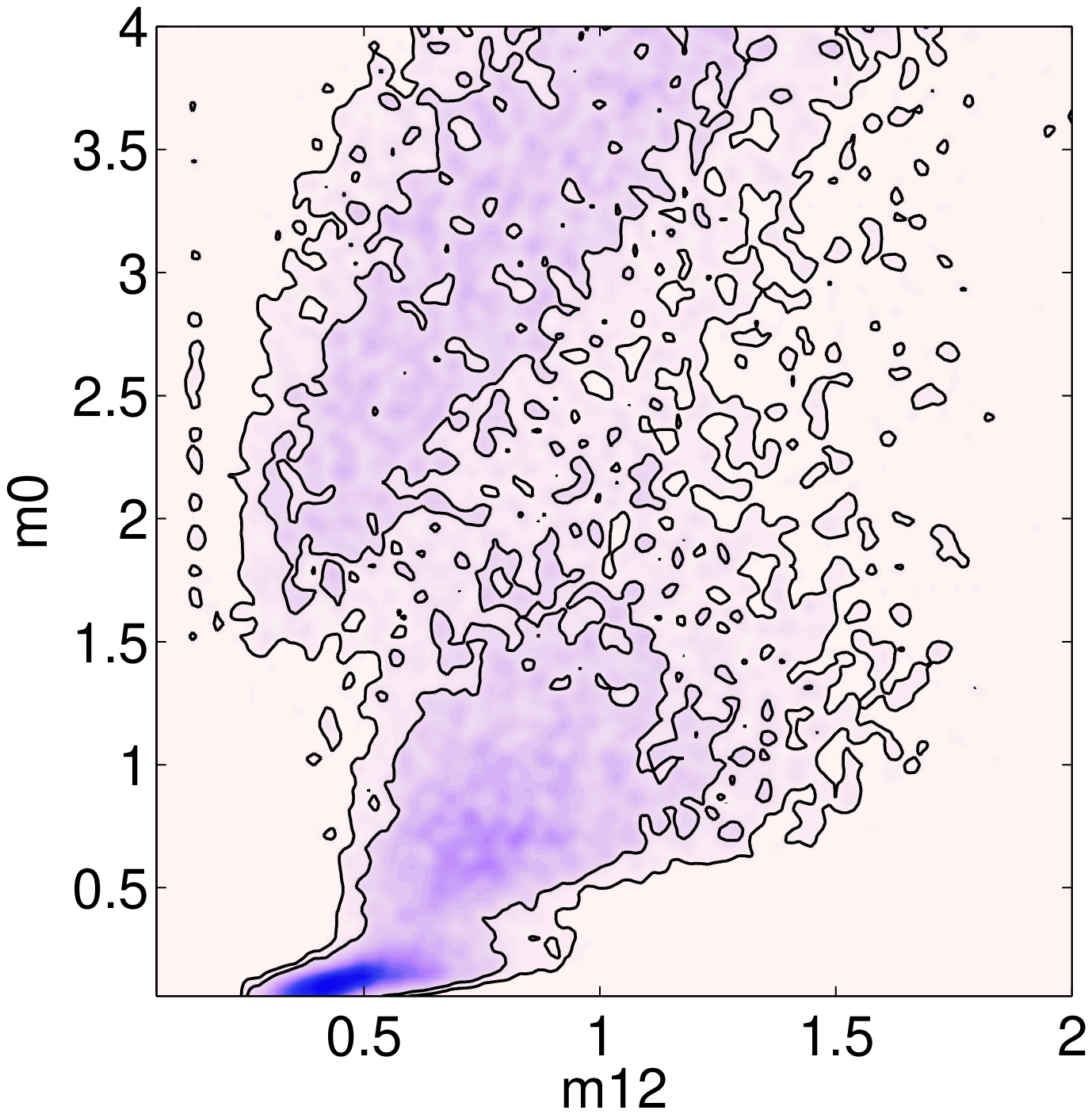}
\includegraphics[width=0.4\columnwidth, height=0.4\columnwidth]{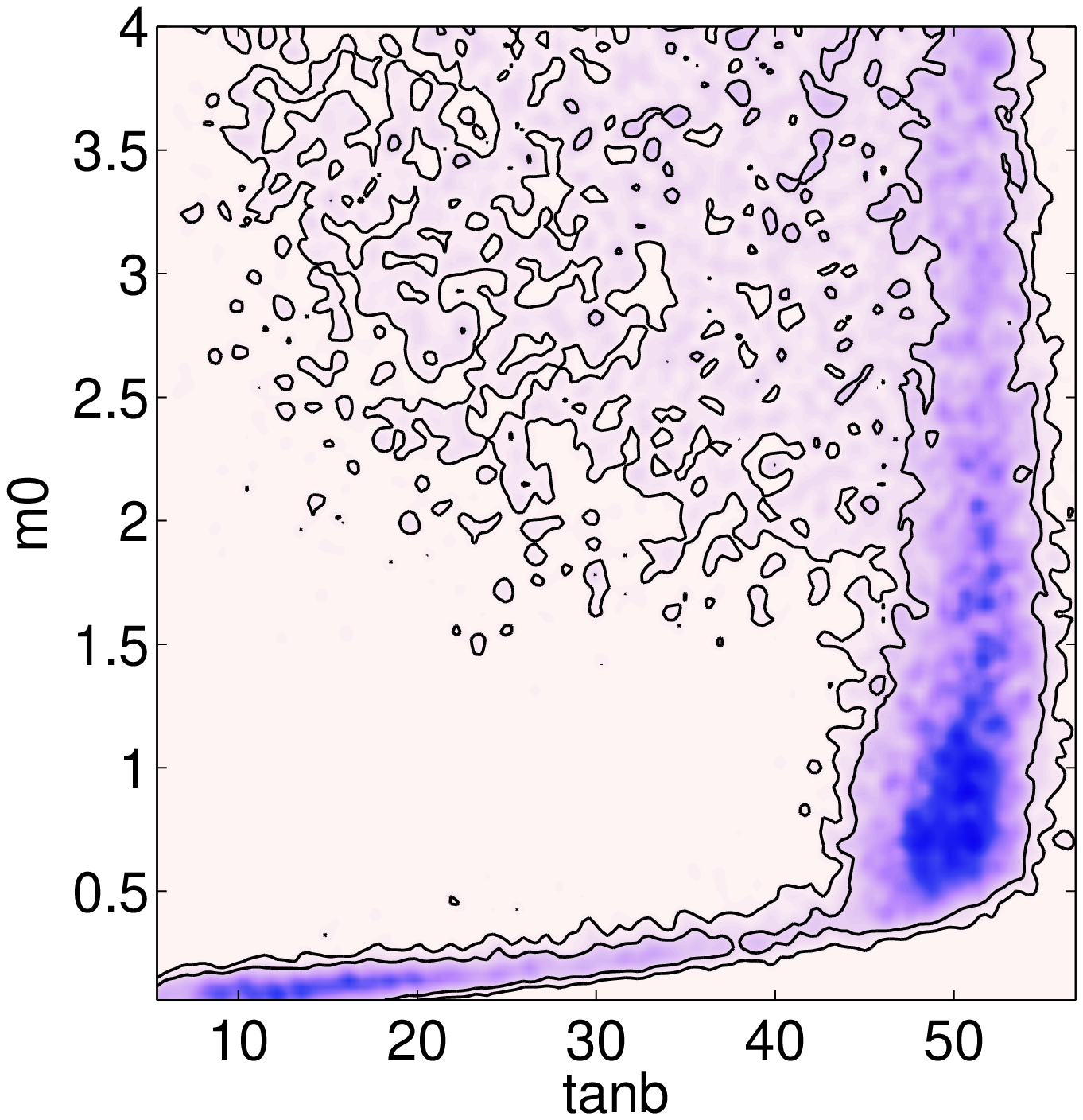}
\includegraphics[width=0.7\columnwidth, height=0.05\columnwidth]{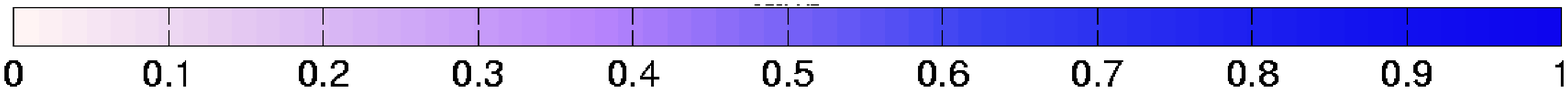}
\caption{The 2-dimensional posterior probability densities in the plane spanned by mSUGRA parameters: $m_0$, 
$m_{1/2}$, $A_0$ and $\tan\beta$ for the linear prior measure ``4 TeV'' range analysis and $\mu > 0$. The inner
and outer contours enclose $68$\% and $95$\% of the total probability respectively. All of the other parameters
in each plane have been marginalised over.}
\label{fig:msugra_2d}}

We display the results of the {\sc MultiNest} fits on the $m_0-M_{1/2}$ and $m_0-\tan \beta$ plane posterior
probability densities in Fig.~\ref{fig:msugra_2d} \footnote{The uneven ``bobbly'' appearance of the 2d
marginalized posteriors is due to the small pixel size used in the marginalized grid; this was required in order
to resolve the finest features in the posterior distributions}. Previous global fits in mSUGRA have found that
the dark matter relic density has the largest effect on parameter space~\cite{Allanach:2005kz}. In particular,
regions where the LSP annihilates efficiently through some particular mechanism are preferred by the fits. In the
left-hand panel, we see that the highest posterior region is where the stau co-annihilation channel is active at
the lowest value of $m_0$, where the lightest stau co-annihilates very efficiently with the lightest neutralino
due to their near mass degeneracy. Next, in the approximate region 0.5 TeV $< m_0 <$ 1.5 TeV, there is another
reasonably high posterior region. In this region, $\tan \beta$ is large and the LSP is approximately half the
mass of the pseudo-scalar Higgs boson $A^0$. The process $\chi_1^0 \chi_1^0 \rightarrow A^0 \rightarrow b {\bar
b}$ becomes an efficient channel in this region. For higher values of $m_0>2$ TeV, the hyperbolic
branch~\cite{Chan:1997bi,feldman08} r\'{e}gime reigns, where the LSP contains a significant higgsino component
and annihilation into weak gauge boson pairs becomes quite efficient. This region dominantly has $\tan \beta >
10$, as can be seen in the right-hand panel of Fig.~\ref{fig:msugra_2d}. All of the qualitative features of
previous MCMC global fits~\cite{Allanach:2005kz,deAustri:2006pe,darkSide,Allanach:2006jc,Allanach:2007} have been
reproduced in the figure, providing a useful validation of the {\sc MultiNest} technique in a particle physics
context, where the shape of the multi-dimensional posterior exhibits multi-modality and curving degeneracies.
2-dimensional marginalisations in other mSUGRA parameter combinations also agree to a large extent with previous
MCMC fits, for both $\mu>0$ and $\mu<0$. However, compared to MCMC fits in
Refs.~\cite{Allanach:2005kz,deAustri:2006pe,rosz:2007}, there has been a slight migration for $\mu>0$: the stau
co-annihilation region has become relatively more favoured than previously and the hyperbolic branch has become
less favoured. This is primarily due to $M_W$ and $\sin^2 \theta_w^l$: our calculation includes 2-loop MSSM
effects and so we are able to place smaller errors on the theoretical prediction than
Refs.~\cite{Allanach:2005kz,deAustri:2006pe,rosz:2007}. Both of these variables show a mild preference for a
sizable SUSY contribution once these 2-loop effects are included~\cite{EWobs}. The pure {\tt SOFTSUSY2.0.17}
calculation is at 1-loop order and without the additional two loop effects, it displays a preference for larger
SUSY scalar masses~\cite{darkSide}, thus favouring the hyperbolic branch region more. An effect in the opposite
direction that comes from including the NNLO corrections to $BR(b \rightarrow s \gamma)$ is
possible~\cite{rosz:2007}. Large values of $m_0$ in the hyperbolic branch region lead to fairly light charged 
Higgs' in mSUGRA due to charged Higgs-top loops, which may then push the branching ratio toward its
experimentally preferred range, by adding constructively to the Standard Model contribution. However, our
estimate of the combined statistical error of $BR(b \rightarrow s \gamma)$ in Table~\ref{tab:observables} means 
that this effect only has a small statistical pull on the fits, being out-weighed by the effects mentioned above 
in the opposite direction. We note here that, as $m_t$ as determined from experiment increases, the focus point
region moves to higher values of $m_0$~\cite{m0mt}. However, very similar fits to the ones presented here were
performed for $m_t=172.6 \pm 1.8$ GeV, see Fig. 2a of Ref.~\cite{Allanach:2008iq}, and the posterior density on
the $m_0-M_{1/2}$ plane did not change much compared to the present paper (which uses $m_t=170.9 \pm 1.8$ GeV). 

\FIGURE{
\psfrag{m12}{$m_{1/2}$ (TeV)}
\psfrag{mhalf}{$m_{1/2}$ (TeV)}
\psfrag{m0}{$m_0$ (TeV)}
\psfrag{A0}{$A_0$ (TeV)}
\psfrag{tanb}{$\tan \beta$}
\includegraphics[width=0.4\columnwidth]{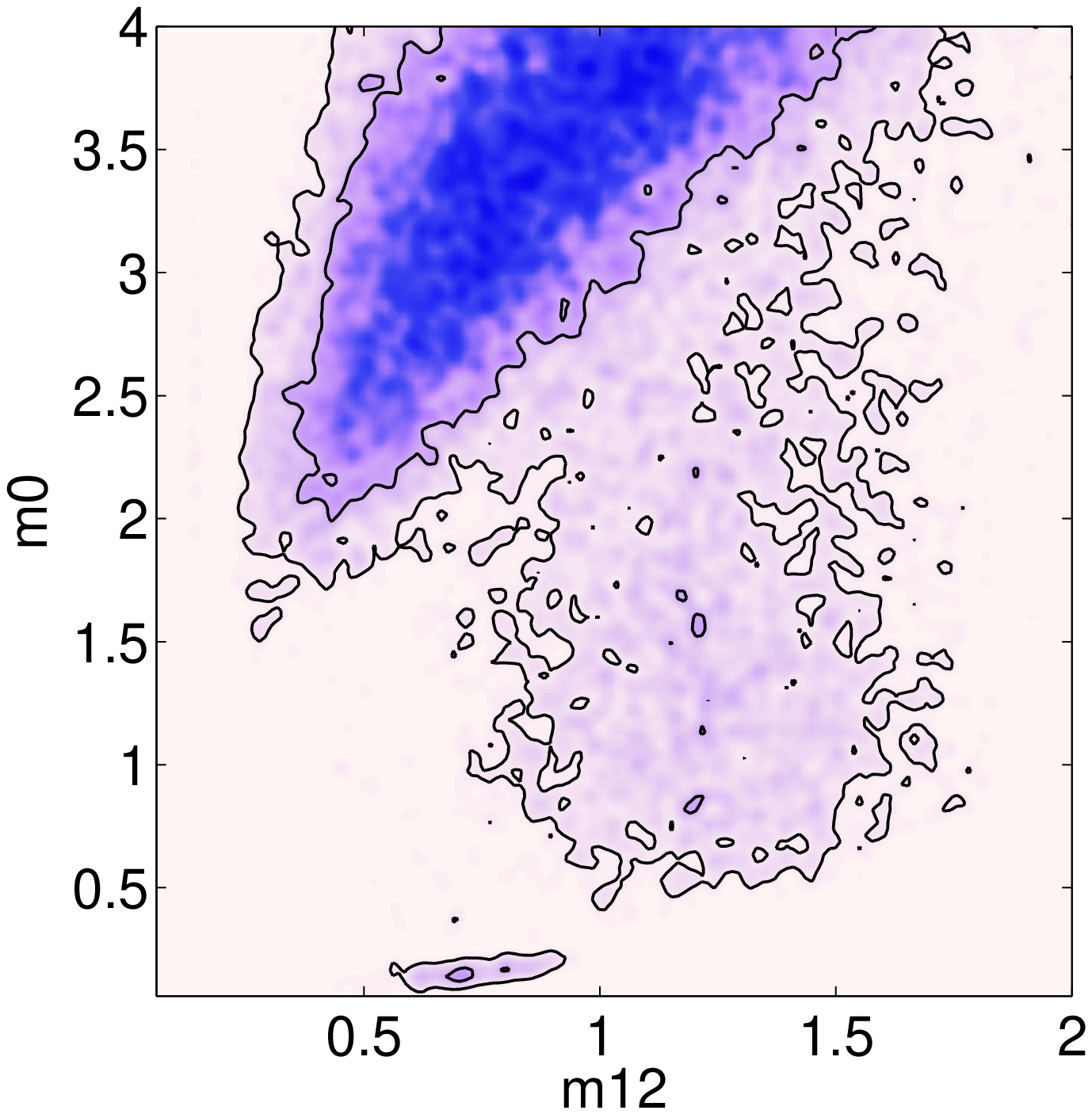}
\includegraphics[width=0.4\columnwidth]{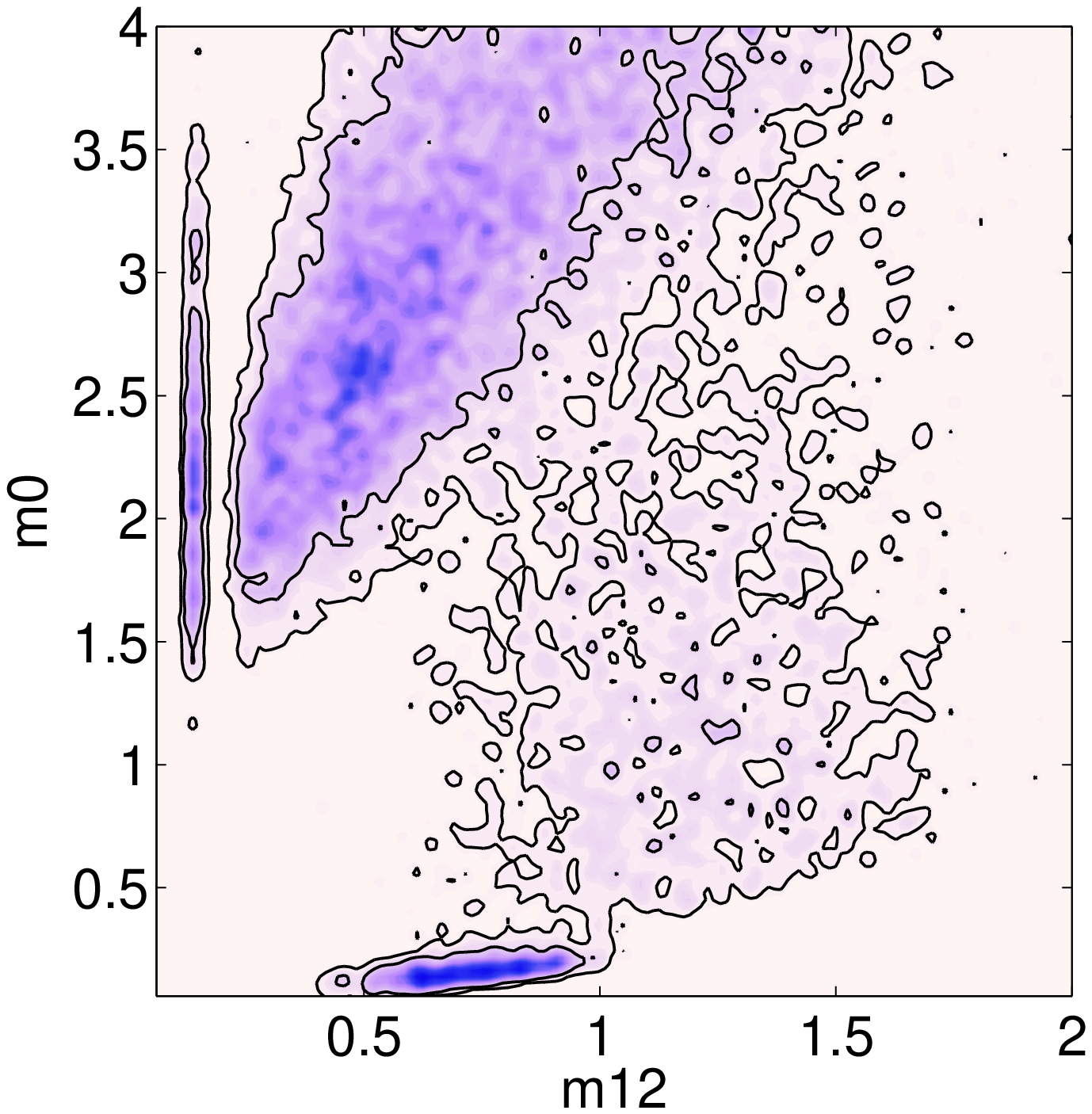}
\includegraphics[width=0.7\columnwidth, height=0.05\columnwidth]{FIGS/colormap.eps}
\caption{The 2-dimensional mSUGRA posterior probability densities in the plane $m_0$, $m_{1/2}$ for $\mu < 0$ for
(left) the `4 TeV range' linear measure prior analysis and (right) the `4 TeV range' logarithmic measure prior
analysis. The inner and outer contours enclose $68$\% and $95$\% of the total probability respectively. All of
the other parameters in each plane have been marginalised over.}
\label{fig:msugra_2d_neg}}

For $\mu<0$, the fit prefers a higher posterior probability for the focus
point region compared to 
Ref.~\cite{darkSide}. We show the marginalisation of $\mu<0$ mSUGRA to the
$m_0-m_{1/2}$ plane in 
Fig.~\ref{fig:msugra_2d_neg}. The left-hand panel shows the linear measure prior analysis and may be compared
directly with Fig. 5a of Ref.~\cite{darkSide}, which has the stau co-annihilation region having the highest
posterior density. The increased discrepancy of $(g-2)_\mu$ in the present fit with current data will favour
heavier sparticles due to the SUSY contribution being of the wrong sign for $\mu<0$ mSUGRA\@. In the right-hand
side, we see how the fit changes due to a logarithmic measure on the prior. Indeed, the foreseen shift toward
lower values of $m_0$ is significant, the stau co-annihilation channel being favoured once more. Although there
are some similarities with the left-hand panel, it is clear that the choice of prior measure still has a
non-negligible effect on the fit despite the inclusion of new $b$-physics observables. 

\FIGURE{
\scalebox{1.0}{
\begingroup
  \makeatletter
  \providecommand\color[2][]{%
    \GenericError{(gnuplot) \space\space\space\@spaces}{%
      Package color not loaded in conjunction with
      terminal option `colourtext'%
    }{See the gnuplot documentation for explanation.%
    }{Either use 'blacktext' in gnuplot or load the package
      color.sty in LaTeX.}%
    \renewcommand\color[2][]{}%
  }%
  \providecommand\includegraphics[2][]{%
    \GenericError{(gnuplot) \space\space\space\@spaces}{%
      Package graphicx or graphics not loaded%
    }{See the gnuplot documentation for explanation.%
    }{The gnuplot epslatex terminal needs graphicx.sty or graphics.sty.}%
    \renewcommand\includegraphics[2][]{}%
  }%
  \providecommand\rotatebox[2]{#2}%
  \@ifundefined{ifGPcolor}{%
    \newif\ifGPcolor
    \GPcolortrue
  }{}%
  \@ifundefined{ifGPblacktext}{%
    \newif\ifGPblacktext
    \GPblacktexttrue
  }{}%
  \let\gplgaddtomacro\g@addto@macro
  \gdef\gplbacktext{}%
  \gdef\gplfronttext{}%
  \makeatother
  \ifGPblacktext
    \def\colorrgb#1{}%
    \def\colorgray#1{}%
  \else
    \ifGPcolor
      \def\colorrgb#1{\color[rgb]{#1}}%
      \def\colorgray#1{\color[gray]{#1}}%
      \expandafter\def\csname LTw\endcsname{\color{white}}%
      \expandafter\def\csname LTb\endcsname{\color{black}}%
      \expandafter\def\csname LTa\endcsname{\color{black}}%
      \expandafter\def\csname LT0\endcsname{\color[rgb]{1,0,0}}%
      \expandafter\def\csname LT1\endcsname{\color[rgb]{0,1,0}}%
      \expandafter\def\csname LT2\endcsname{\color[rgb]{0,0,1}}%
      \expandafter\def\csname LT3\endcsname{\color[rgb]{1,0,1}}%
      \expandafter\def\csname LT4\endcsname{\color[rgb]{0,1,1}}%
      \expandafter\def\csname LT5\endcsname{\color[rgb]{1,1,0}}%
      \expandafter\def\csname LT6\endcsname{\color[rgb]{0,0,0}}%
      \expandafter\def\csname LT7\endcsname{\color[rgb]{1,0.3,0}}%
      \expandafter\def\csname LT8\endcsname{\color[rgb]{0.5,0.5,0.5}}%
    \else
      \def\colorrgb#1{\color{black}}%
      \def\colorgray#1{\color[gray]{#1}}%
      \expandafter\def\csname LTw\endcsname{\color{white}}%
      \expandafter\def\csname LTb\endcsname{\color{black}}%
      \expandafter\def\csname LTa\endcsname{\color{black}}%
      \expandafter\def\csname LT0\endcsname{\color{black}}%
      \expandafter\def\csname LT1\endcsname{\color{black}}%
      \expandafter\def\csname LT2\endcsname{\color{black}}%
      \expandafter\def\csname LT3\endcsname{\color{black}}%
      \expandafter\def\csname LT4\endcsname{\color{black}}%
      \expandafter\def\csname LT5\endcsname{\color{black}}%
      \expandafter\def\csname LT6\endcsname{\color{black}}%
      \expandafter\def\csname LT7\endcsname{\color{black}}%
      \expandafter\def\csname LT8\endcsname{\color{black}}%
    \fi
  \fi
  \setlength{\unitlength}{0.0500bp}%
  \begin{picture}(7200.00,5040.00)%
    \gplgaddtomacro\gplbacktext{%
      \csname LTb\endcsname%
      \put(1122,3180){\makebox(0,0)[r]{\strut{} 0}}%
      \put(1122,3579){\makebox(0,0)[r]{\strut{} 0.25}}%
      \put(1122,3978){\makebox(0,0)[r]{\strut{} 0.5}}%
      \put(1122,4377){\makebox(0,0)[r]{\strut{} 0.75}}%
      \put(1122,4776){\makebox(0,0)[r]{\strut{} 1}}%
      \put(1254,2960){\makebox(0,0){\strut{} 0}}%
      \put(1747,2960){\makebox(0,0){\strut{} 1}}%
      \put(2240,2960){\makebox(0,0){\strut{} 2}}%
      \put(2733,2960){\makebox(0,0){\strut{} 3}}%
      \put(3226,2960){\makebox(0,0){\strut{} 4}}%
      \put(220,3978){\rotatebox{90}{\makebox(0,0){\strut{}$P/P_{max}$}}}%
      \put(2240,2630){\makebox(0,0){\strut{}$m_0$ (TeV)}}%
    }%
    \gplgaddtomacro\gplfronttext{%
    }%
    \gplgaddtomacro\gplbacktext{%
      \csname LTb\endcsname%
      \put(4502,3180){\makebox(0,0)[r]{\strut{} 0}}%
      \put(4502,3579){\makebox(0,0)[r]{\strut{} 0.25}}%
      \put(4502,3978){\makebox(0,0)[r]{\strut{} 0.5}}%
      \put(4502,4377){\makebox(0,0)[r]{\strut{} 0.75}}%
      \put(4502,4776){\makebox(0,0)[r]{\strut{} 1}}%
      \put(4634,2960){\makebox(0,0){\strut{} 0}}%
      \put(5182,2960){\makebox(0,0){\strut{} 0.5}}%
      \put(5730,2960){\makebox(0,0){\strut{} 1}}%
      \put(6278,2960){\makebox(0,0){\strut{} 1.5}}%
      \put(6826,2960){\makebox(0,0){\strut{} 2}}%
      \put(3820,3978){\rotatebox{90}{\makebox(0,0){\strut{}}}}%
      \put(5730,2630){\makebox(0,0){\strut{}$m_{1/2}$ (TeV)}}%
    }%
    \gplgaddtomacro\gplfronttext{%
      \csname LTb\endcsname%
      \put(5839,4603){\makebox(0,0)[r]{\strut{}$\mu > 0$}}%
      \csname LTb\endcsname%
      \put(5839,4383){\makebox(0,0)[r]{\strut{}$\mu < 0$}}%
    }%
    \gplgaddtomacro\gplbacktext{%
      \csname LTb\endcsname%
      \put(1122,660){\makebox(0,0)[r]{\strut{} 0}}%
      \put(1122,1059){\makebox(0,0)[r]{\strut{} 0.25}}%
      \put(1122,1458){\makebox(0,0)[r]{\strut{} 0.5}}%
      \put(1122,1857){\makebox(0,0)[r]{\strut{} 0.75}}%
      \put(1122,2256){\makebox(0,0)[r]{\strut{} 1}}%
      \put(1254,440){\makebox(0,0){\strut{}-8}}%
      \put(1536,440){\makebox(0,0){\strut{}-6}}%
      \put(1817,440){\makebox(0,0){\strut{}-4}}%
      \put(2099,440){\makebox(0,0){\strut{}-2}}%
      \put(2381,440){\makebox(0,0){\strut{} 0}}%
      \put(2663,440){\makebox(0,0){\strut{} 2}}%
      \put(2944,440){\makebox(0,0){\strut{} 4}}%
      \put(3226,440){\makebox(0,0){\strut{} 6}}%
      \put(220,1458){\rotatebox{90}{\makebox(0,0){\strut{}$P/P_{max}$}}}%
      \put(2240,110){\makebox(0,0){\strut{}$A_0$ (TeV)}}%
    }%
    \gplgaddtomacro\gplfronttext{%
    }%
    \gplgaddtomacro\gplbacktext{%
      \csname LTb\endcsname%
      \put(4502,660){\makebox(0,0)[r]{\strut{} 0}}%
      \put(4502,1059){\makebox(0,0)[r]{\strut{} 0.25}}%
      \put(4502,1458){\makebox(0,0)[r]{\strut{} 0.5}}%
      \put(4502,1857){\makebox(0,0)[r]{\strut{} 0.75}}%
      \put(4502,2256){\makebox(0,0)[r]{\strut{} 1}}%
      \put(4634,440){\makebox(0,0){\strut{} 0}}%
      \put(4999,440){\makebox(0,0){\strut{} 10}}%
      \put(5365,440){\makebox(0,0){\strut{} 20}}%
      \put(5730,440){\makebox(0,0){\strut{} 30}}%
      \put(6095,440){\makebox(0,0){\strut{} 40}}%
      \put(6461,440){\makebox(0,0){\strut{} 50}}%
      \put(6826,440){\makebox(0,0){\strut{} 60}}%
      \put(3820,1458){\rotatebox{90}{\makebox(0,0){\strut{}}}}%
      \put(5730,110){\makebox(0,0){\strut{}$\tan\beta$}}%
    }%
    \gplgaddtomacro\gplfronttext{%
    }%
    \gplbacktext
    \put(0,0){\includegraphics{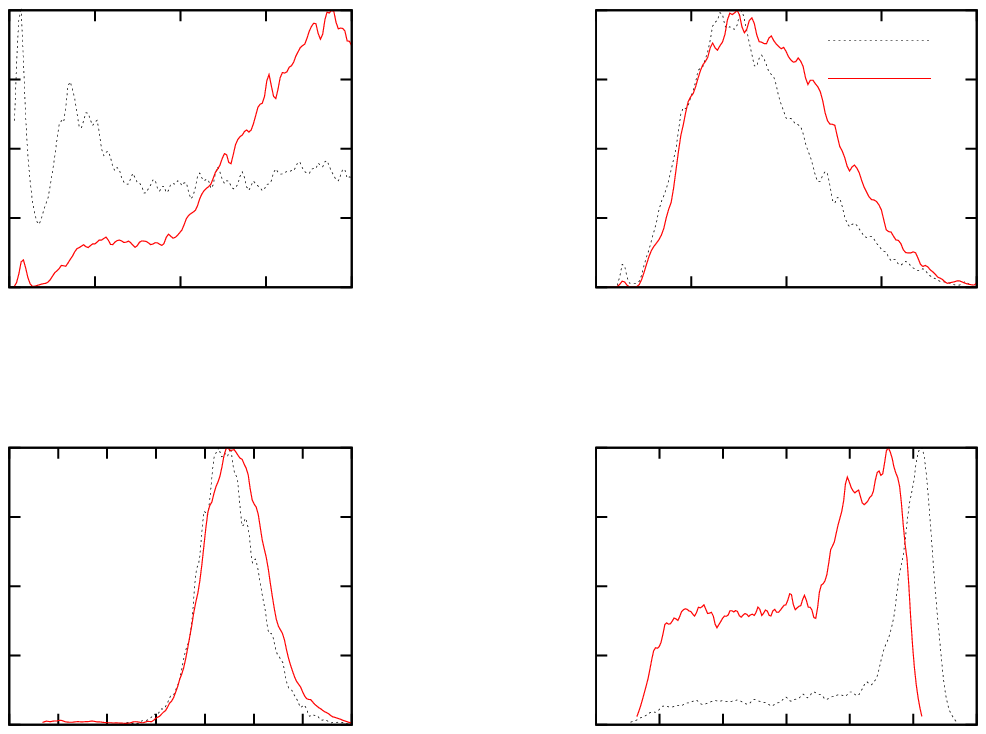}}%
    \gplfronttext
  \end{picture}%
\endgroup
}
\caption{Comparison of $\mu<0$ and $\mu>0$ 1-dimensional relative posterior probability densities of mSUGRA
 parameters for the linear measure prior `4 TeV' range analysis. All of the other input parameters have been
 marginalised over. \label{fig:msugra_1d}}}

With this fact still in mind, we compare the posterior probability density function for $\mu>0$ and $\mu<0$ in
Fig.~\ref{fig:msugra_1d} for linear measure priors. In Fig.~\ref{fig:msugra_1d}, we see the preference for
heavier sparticles in the $\mu<0$ case reflected in the larger values for the universal scalar and gaugino masses
$m_0$ and $m_{1/2}$. It is clear from the top left hand panel that any inference made about scalar masses $\mu<0$
will be quite sensitive to the exact range taken, since the $m_0$ distribution is near its maximum at large
values close to 4 TeV. On the other hand, the data constrains $m_{1/2} < 2$ TeV robustly. $\mu<0$ favours large
$\tan \beta$ less than $\mu>0$ since for large $\tan \beta$, $(g-2)_\mu$ becomes more negative, with the wrong
sign compared to the data. 

\FIGURE{
\scalebox{0.8}{
\begingroup
  \makeatletter
  \providecommand\color[2][]{%
    \GenericError{(gnuplot) \space\space\space\@spaces}{%
      Package color not loaded in conjunction with
      terminal option `colourtext'%
    }{See the gnuplot documentation for explanation.%
    }{Either use 'blacktext' in gnuplot or load the package
      color.sty in LaTeX.}%
    \renewcommand\color[2][]{}%
  }%
  \providecommand\includegraphics[2][]{%
    \GenericError{(gnuplot) \space\space\space\@spaces}{%
      Package graphicx or graphics not loaded%
    }{See the gnuplot documentation for explanation.%
    }{The gnuplot epslatex terminal needs graphicx.sty or graphics.sty.}%
    \renewcommand\includegraphics[2][]{}%
  }%
  \providecommand\rotatebox[2]{#2}%
  \@ifundefined{ifGPcolor}{%
    \newif\ifGPcolor
    \GPcolortrue
  }{}%
  \@ifundefined{ifGPblacktext}{%
    \newif\ifGPblacktext
    \GPblacktexttrue
  }{}%
  \let\gplgaddtomacro\g@addto@macro
  \gdef\gplbacktext{}%
  \gdef\gplfronttext{}%
  \makeatother
  \ifGPblacktext
    \def\colorrgb#1{}%
    \def\colorgray#1{}%
  \else
    \ifGPcolor
      \def\colorrgb#1{\color[rgb]{#1}}%
      \def\colorgray#1{\color[gray]{#1}}%
      \expandafter\def\csname LTw\endcsname{\color{white}}%
      \expandafter\def\csname LTb\endcsname{\color{black}}%
      \expandafter\def\csname LTa\endcsname{\color{black}}%
      \expandafter\def\csname LT0\endcsname{\color[rgb]{1,0,0}}%
      \expandafter\def\csname LT1\endcsname{\color[rgb]{0,1,0}}%
      \expandafter\def\csname LT2\endcsname{\color[rgb]{0,0,1}}%
      \expandafter\def\csname LT3\endcsname{\color[rgb]{1,0,1}}%
      \expandafter\def\csname LT4\endcsname{\color[rgb]{0,1,1}}%
      \expandafter\def\csname LT5\endcsname{\color[rgb]{1,1,0}}%
      \expandafter\def\csname LT6\endcsname{\color[rgb]{0,0,0}}%
      \expandafter\def\csname LT7\endcsname{\color[rgb]{1,0.3,0}}%
      \expandafter\def\csname LT8\endcsname{\color[rgb]{0.5,0.5,0.5}}%
    \else
      \def\colorrgb#1{\color{black}}%
      \def\colorgray#1{\color[gray]{#1}}%
      \expandafter\def\csname LTw\endcsname{\color{white}}%
      \expandafter\def\csname LTb\endcsname{\color{black}}%
      \expandafter\def\csname LTa\endcsname{\color{black}}%
      \expandafter\def\csname LT0\endcsname{\color{black}}%
      \expandafter\def\csname LT1\endcsname{\color{black}}%
      \expandafter\def\csname LT2\endcsname{\color{black}}%
      \expandafter\def\csname LT3\endcsname{\color{black}}%
      \expandafter\def\csname LT4\endcsname{\color{black}}%
      \expandafter\def\csname LT5\endcsname{\color{black}}%
      \expandafter\def\csname LT6\endcsname{\color{black}}%
      \expandafter\def\csname LT7\endcsname{\color{black}}%
      \expandafter\def\csname LT8\endcsname{\color{black}}%
    \fi
  \fi
  \setlength{\unitlength}{0.0500bp}%
  \begin{picture}(10800.00,7560.00)%
    \gplgaddtomacro\gplbacktext{%
      \csname LTb\endcsname%
      \put(1122,5700){\makebox(0,0)[r]{\strut{} 0}}%
      \put(1122,6099){\makebox(0,0)[r]{\strut{} 0.25}}%
      \put(1122,6498){\makebox(0,0)[r]{\strut{} 0.5}}%
      \put(1122,6897){\makebox(0,0)[r]{\strut{} 0.75}}%
      \put(1122,7296){\makebox(0,0)[r]{\strut{} 1}}%
      \put(1536,5480){\makebox(0,0){\strut{} 0.0343}}%
      \put(2193,5480){\makebox(0,0){\strut{} 0.1143}}%
      \put(2850,5480){\makebox(0,0){\strut{} 0.1943}}%
      \put(220,6498){\rotatebox{90}{\makebox(0,0){\strut{}$P/P_{max}$}}}%
      \put(2240,5150){\makebox(0,0){\strut{}$\Omega_{DM} h^2$}}%
    }%
    \gplgaddtomacro\gplfronttext{%
    }%
    \gplgaddtomacro\gplbacktext{%
      \csname LTb\endcsname%
      \put(4502,5700){\makebox(0,0)[r]{\strut{} 0}}%
      \put(4502,6099){\makebox(0,0)[r]{\strut{} 0.25}}%
      \put(4502,6498){\makebox(0,0)[r]{\strut{} 0.5}}%
      \put(4502,6897){\makebox(0,0)[r]{\strut{} 0.75}}%
      \put(4502,7296){\makebox(0,0)[r]{\strut{} 1}}%
      \put(4908,5480){\makebox(0,0){\strut{} 3.1}}%
      \put(5730,5480){\makebox(0,0){\strut{} 29.5}}%
      \put(6552,5480){\makebox(0,0){\strut{} 55.9}}%
      \put(3820,6498){\rotatebox{90}{\makebox(0,0){\strut{}}}}%
      \put(5730,5150){\makebox(0,0){\strut{}$\delta a_{\mu} \times 10^{10}$}}%
    }%
    \gplgaddtomacro\gplfronttext{%
      \csname LTb\endcsname%
      \put(6354,7145){\makebox(0,0)[r]{\strut{}$\mu > 0$}}%
      \csname LTb\endcsname%
      \put(6354,6969){\makebox(0,0)[r]{\strut{}$\mu < 0$}}%
      \csname LTb\endcsname%
      \put(6354,6793){\makebox(0,0)[r]{\strut{}likelihood}}%
    }%
    \gplgaddtomacro\gplbacktext{%
      \csname LTb\endcsname%
      \put(8102,5700){\makebox(0,0)[r]{\strut{} 0}}%
      \put(8102,6099){\makebox(0,0)[r]{\strut{} 0.25}}%
      \put(8102,6498){\makebox(0,0)[r]{\strut{} 0.5}}%
      \put(8102,6897){\makebox(0,0)[r]{\strut{} 0.75}}%
      \put(8102,7296){\makebox(0,0)[r]{\strut{} 1}}%
      \put(8508,5480){\makebox(0,0){\strut{} 1.39}}%
      \put(9330,5480){\makebox(0,0){\strut{} 3.55}}%
      \put(10152,5480){\makebox(0,0){\strut{} 5.71}}%
      \put(7420,6498){\rotatebox{90}{\makebox(0,0){\strut{}}}}%
      \put(9330,5150){\makebox(0,0){\strut{}$BR(b \rightarrow s \gamma) \times 10^4$}}%
    }%
    \gplgaddtomacro\gplfronttext{%
    }%
    \gplgaddtomacro\gplbacktext{%
      \csname LTb\endcsname%
      \put(1122,3180){\makebox(0,0)[r]{\strut{} 0}}%
      \put(1122,3579){\makebox(0,0)[r]{\strut{} 0.25}}%
      \put(1122,3978){\makebox(0,0)[r]{\strut{} 0.5}}%
      \put(1122,4377){\makebox(0,0)[r]{\strut{} 0.75}}%
      \put(1122,4776){\makebox(0,0)[r]{\strut{} 1}}%
      \put(1501,2960){\makebox(0,0){\strut{} 80.317}}%
      \put(2240,2960){\makebox(0,0){\strut{} 80.398}}%
      \put(2980,2960){\makebox(0,0){\strut{} 80.479}}%
      \put(220,3978){\rotatebox{90}{\makebox(0,0){\strut{}$P/P_{max}$}}}%
      \put(2240,2630){\makebox(0,0){\strut{}$m_W$}}%
    }%
    \gplgaddtomacro\gplfronttext{%
    }%
    \gplgaddtomacro\gplbacktext{%
      \csname LTb\endcsname%
      \put(4502,3180){\makebox(0,0)[r]{\strut{} 0}}%
      \put(4502,3579){\makebox(0,0)[r]{\strut{} 0.25}}%
      \put(4502,3978){\makebox(0,0)[r]{\strut{} 0.5}}%
      \put(4502,4377){\makebox(0,0)[r]{\strut{} 0.75}}%
      \put(4502,4776){\makebox(0,0)[r]{\strut{} 1}}%
      \put(4908,2960){\makebox(0,0){\strut{} 0.230971}}%
      \put(5730,2960){\makebox(0,0){\strut{} 0.23149}}%
      \put(6552,2960){\makebox(0,0){\strut{} 0.232009}}%
      \put(3820,3978){\rotatebox{90}{\makebox(0,0){\strut{}}}}%
      \put(5730,2630){\makebox(0,0){\strut{}$\sin^2 \theta_w^l$}}%
    }%
    \gplgaddtomacro\gplfronttext{%
    }%
    \gplgaddtomacro\gplbacktext{%
      \csname LTb\endcsname%
      \put(8102,3180){\makebox(0,0)[r]{\strut{} 0}}%
      \put(8102,3579){\makebox(0,0)[r]{\strut{} 0.25}}%
      \put(8102,3978){\makebox(0,0)[r]{\strut{} 0.5}}%
      \put(8102,4377){\makebox(0,0)[r]{\strut{} 0.75}}%
      \put(8102,4776){\makebox(0,0)[r]{\strut{} 1}}%
      \put(8508,2960){\makebox(0,0){\strut{} 0.125}}%
      \put(9330,2960){\makebox(0,0){\strut{} 1.259}}%
      \put(10152,2960){\makebox(0,0){\strut{} 2.393}}%
      \put(7420,3978){\rotatebox{90}{\makebox(0,0){\strut{}}}}%
      \put(9330,2630){\makebox(0,0){\strut{}$R_{BR(B_u \rightarrow \tau \nu)}$}}%
    }%
    \gplgaddtomacro\gplfronttext{%
    }%
    \gplgaddtomacro\gplbacktext{%
      \csname LTb\endcsname%
      \put(2922,660){\makebox(0,0)[r]{\strut{} 0}}%
      \put(2922,1059){\makebox(0,0)[r]{\strut{} 0.25}}%
      \put(2922,1458){\makebox(0,0)[r]{\strut{} 0.5}}%
      \put(2922,1857){\makebox(0,0)[r]{\strut{} 0.75}}%
      \put(2922,2256){\makebox(0,0)[r]{\strut{} 1}}%
      \put(3301,440){\makebox(0,0){\strut{}-0.0492}}%
      \put(4040,440){\makebox(0,0){\strut{} 0.0375}}%
      \put(4780,440){\makebox(0,0){\strut{} 0.1242}}%
      \put(2020,1458){\rotatebox{90}{\makebox(0,0){\strut{}$P/P_{max}$}}}%
      \put(4040,110){\makebox(0,0){\strut{}$\Delta_{o-}$}}%
    }%
    \gplgaddtomacro\gplfronttext{%
    }%
    \gplgaddtomacro\gplbacktext{%
      \csname LTb\endcsname%
      \put(6302,660){\makebox(0,0)[r]{\strut{} 0}}%
      \put(6302,1059){\makebox(0,0)[r]{\strut{} 0.25}}%
      \put(6302,1458){\makebox(0,0)[r]{\strut{} 0.5}}%
      \put(6302,1857){\makebox(0,0)[r]{\strut{} 0.75}}%
      \put(6302,2256){\makebox(0,0)[r]{\strut{} 1}}%
      \put(6708,440){\makebox(0,0){\strut{} 0.49}}%
      \put(7530,440){\makebox(0,0){\strut{} 0.85}}%
      \put(8352,440){\makebox(0,0){\strut{} 1.21}}%
      \put(5620,1458){\rotatebox{90}{\makebox(0,0){\strut{}}}}%
      \put(7530,110){\makebox(0,0){\strut{}$R_{\Delta m_s}$}}%
    }%
    \gplgaddtomacro\gplfronttext{%
    }%
    \gplbacktext
    \put(0,0){\includegraphics{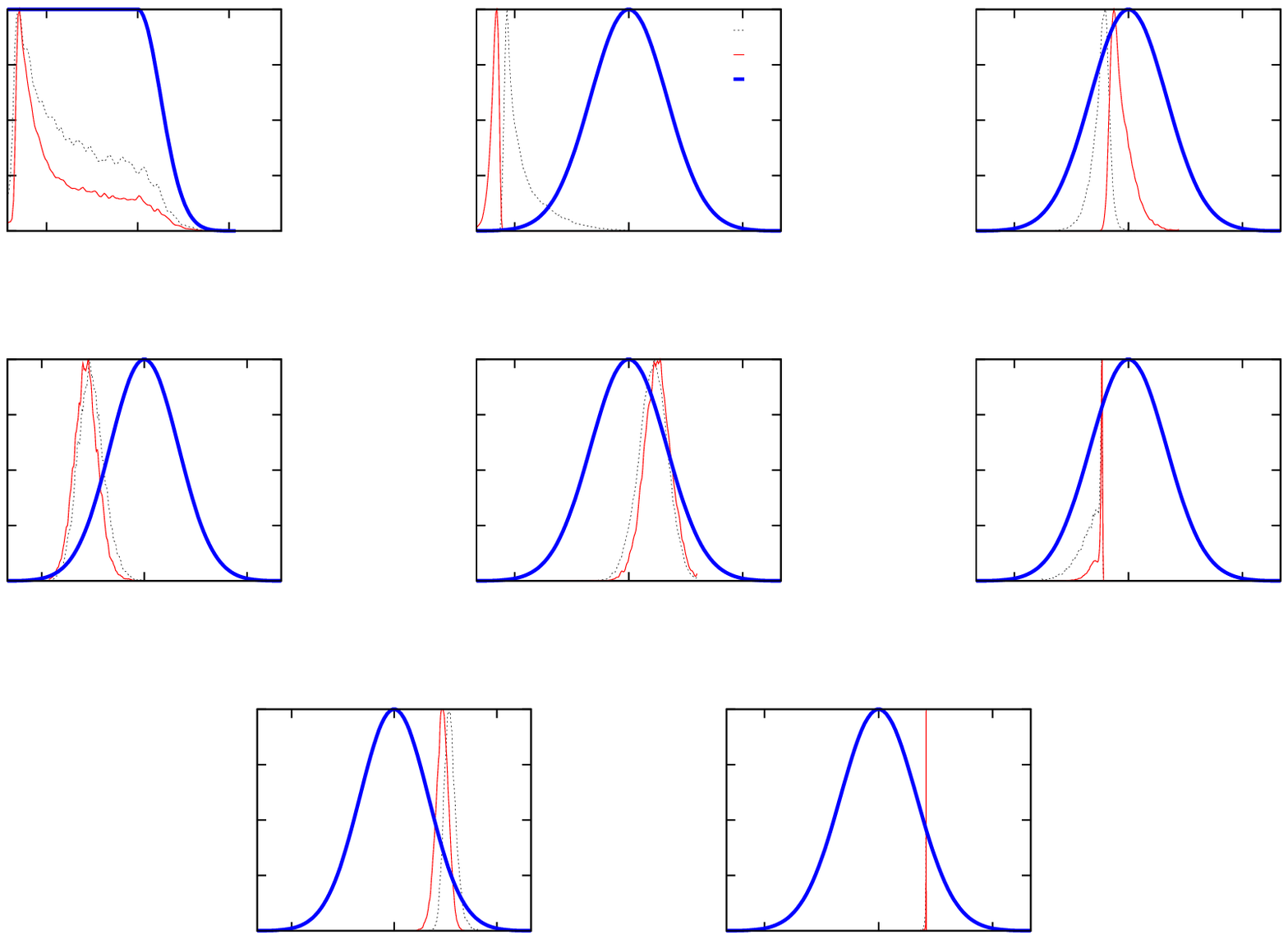}}%
    \gplfronttext
  \end{picture}%
\endgroup
}
\caption{An illustration of tensions between different observables for the mSUGRA model. 
The black (dash-dotted), red (thin solid) and the blue (thick solid) lines show the relative posterior
probability for $\mu > 0$, $\mu < 0$ and the likelihood respectively for each observable.}
\label{fig:cons_1d}}

\TABULAR{|l|cc|cc|}{\hline
            & \multicolumn{2}{c|}{$\mu > 0$}    & \multicolumn{2}{c|}{$\mu < 0$}   \\
Parameter   & $68$\% region   & $95$\% region   & $68$\% region   & $95$\% region  \\ \hline
$m_{h^0}$ (GeV) & $(117,119)$ & $(114,121)$ & $(119,120)$ & $(117,121)$        \\ \hline
$m_{A^0}$ (TeV) & $(0.62,2.12)$ & $(0.48,3.33)$ & $(1.08,3.23)$ & $(0.75,3.75)$  \\ \hline
$m_{\tilde{q}_L}$ (TeV) & $(1.57,3.79)$ & $(0.93,4.47)$ & $(2.71,4.18)$ & $(2.07,4.64)$ \\ \hline
$m_{\tilde{g}}$ (TeV) & $(1.53,2.17)$ & $(0.95,3.15)$ & $(1.75,2.45)$ & $(1.11,3.29)$ \\ \hline
$m_{\tilde{\chi}_1^0}$ (TeV) & $(0.19,0.48)$ & $(0.11,0.68)$ & $(0.20,0.52)$ & $(0.13,0.70)$ \\ \hline
$m_{\tilde{\chi}_1^\pm}$ (TeV) & $(0.25,0.86)$ & $(0.14,1.22)$ & $(0.22,0.88)$ & $(0.15,1.26)$ \\ \hline
$m_{\tilde{e}_R}$ (TeV) & $(0.69,3.34)$ & $(0.21,3.91)$ & $(2.09,3.75)$ & $(0.93,3.97)$ \\ \hline
}
{sparticle mass ranges for linear `4 TeV' analysis corresponding to $68$\% and $95$\% of posterior probability. 
\label{tab:mass_ranges}}

As discussed in Section~\ref{sec:bayesian}, one can easily obtain the posterior for the observables, which are
derived from the model parameters, from the posterior of the model parameters. Fig.~\ref{fig:cons_1d} displays
the statistical pulls of the various observables. In the absence of any tension between the constraints or volume
effects, one would expect the posterior curves to lie on top of the  likelihood curves representing the
experimental data used in the analysis (see also \cite{roszkowski07}). In  order to separate the volume effects
from pulls originating from data, the likelihood profile could be used~\cite{Allanach:2007}. Here though, we just
comment on the combined effect from the two mechanisms. We see that $\Omega_{DM} h^2$ has a preference for being
rather small, but non-zero for either sign of $\mu$. Since any value below $\Omega_{DM} h^2=0.1143$ is not
penalised by the likelihood penalty we have used, this may be ascribed to a combination of volume effects (where
there is simply more volume of parameter space with a small relic density) and pull toward those region from the
other observables. The biggest disparity between the experimental data and the posterior probability distribution
is observed for the $\delta a_\mu$ constraint, which can only be near its central measured value for light
sparticles and large $\tan \beta$. Many of the other constraints are pulling toward large values of the masses,
where the volume of parameter space is larger, and so small values of $|\delta a_\mu|$ are preferred. We see a
slight preference for $\mu<0$ from the $BR(b \rightarrow s \gamma)$ constraint, as expected from the discussion
in Section\ref{sec:const} and Ref.~\cite{rosz:2007}, but this is too small to outweigh the effects of $\delta
a_\mu$, as shown previously by our estimate of $P_+/P_-$. The figure shows that the ratio $R_{\Delta m_s}$, of
the MSSM prediction of the $B_s$ mass splitting to the SM prediction is really not active, i.e.\ that it does not
vary across allowed mSUGRA parameter space, and so does not have an effect on the posterior density.

We list the sparticle mass ranges for linear `4 TeV' analysis corresponding to $68\%$ and $95\%$ of the posterior
probability in Table~\ref{tab:mass_ranges}.

\subsection{Consistency Check between Different Constraints}
\label{sec:results:consistency}

It is clear from Fig.~\ref{fig:cons_1d} that $\delta a_\mu$ and $BR(b \rightarrow s \gamma)$, both important
observables, are pulling in opposite directions. We choose the `strongly preferred' value of $\mu>0$ for our
analysis. In order to check whether the observables $(g-2)_\mu$ and $BR(b\rightarrow s \gamma)$ provide
consistent information on the $\mu > 0$ branch of mSUGRA parameter space, calculation of the parameter $R$ as
given in Eq.~\ref{eq:consistency} is required. In order to carry out this calculation, we impose linear `4 TeV'
priors. In Fig.~\ref{fig:msugra_consistency_own}, we plot the posterior probability distributions for the
$m_0-m_{1/2}$ and $m_0-\tan\beta$ planes for the analysis with $\Omega_{\mathrm{DM}}h^2$, $(g-2)_\mu$ and $BR(b
\rightarrow s \gamma)$ individually. From the figure, we see that the $68$\% probability regions preferred by
the $\delta a_\mu$ and $BR(b \rightarrow s \gamma)$ data are a little different as expected for $\mu>0$, since
$\delta a_\mu$ prefers light SUSY particles whereas the $BR(b \rightarrow s \gamma)$ datum prefers heavy ones in
the hyperbolic branch region. Nevertheless, some overlap in the $95$\% probability regions favoured by these
two data-sets. One would then expect the inconsistency between $BR(b \rightarrow s \gamma)$ and $(g-2)_\mu$ not
to be highly significant. We evaluate 
\begin{equation}
\log R=-0.32 \pm 0.04,
\end{equation}
showing very small evidence for inconsistency between $(g-2)_\mu$ and $BR(b \rightarrow s \gamma)$.

Since $\Omega_{DM} h^2$ plays such a dominant role in shaping the posterior, we next check consistency between
all three constraints in mSUGRA. We perform the analysis in the same manner as described above and evaluate $R$
to be: 
\begin{equation}
\log R=0.61 \pm 0.06,
\end{equation}
showing no evidence for inconsistency between $(g-2)_\mu$, $BR(b \rightarrow s \gamma)$ and
$\Omega_{\mathrm{DM}}h^2$.

These results can be seen qualitatively in the 2-D posterior for the joint analysis of $(g-2)_\mu$,
$BR(b\rightarrow s \gamma)$ and $\Omega_{DM} h^2$ in Fig.~\ref{fig:msugra_consistency_own}. It can be seen that
the joint posterior lies precisely in the region of overlap between posteriors for the analysis of these three
data-sets separately. As shown in Appendix~\ref{app:cosistency}, in the presence of any inconsistency between
different data-sets, the joint posterior can be seen to exclude the high posterior probability regions for the
analysis with the data-sets separately which is not the case here and consequently we do not find a strong
evidence for inconsistency between $(g-2)_\mu$, $BR(b\rightarrow s \gamma)$ and $\Omega_{DM} h^2$ data-sets.

\FIGURE{
\psfrag{m12}{$m_{1/2}$ (TeV)}
\psfrag{m0}{$m_0$ (TeV)}
\psfrag{tanb}{$\tan \beta$}
{\includegraphics[width=0.29\columnwidth,height=0.29\columnwidth]{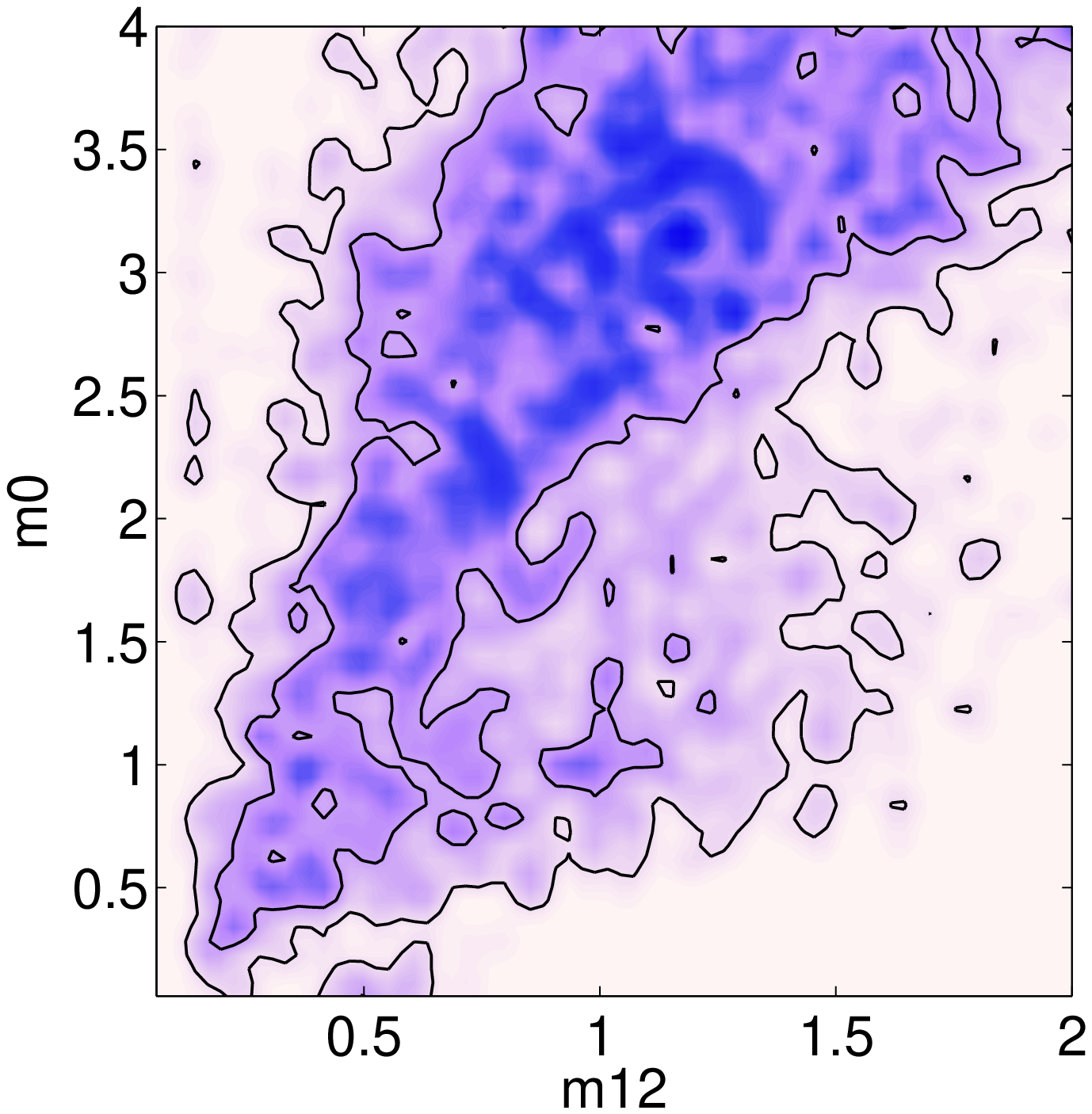}}\hspace{0.3cm}
{\includegraphics[width=0.29\columnwidth,height=0.29\columnwidth]{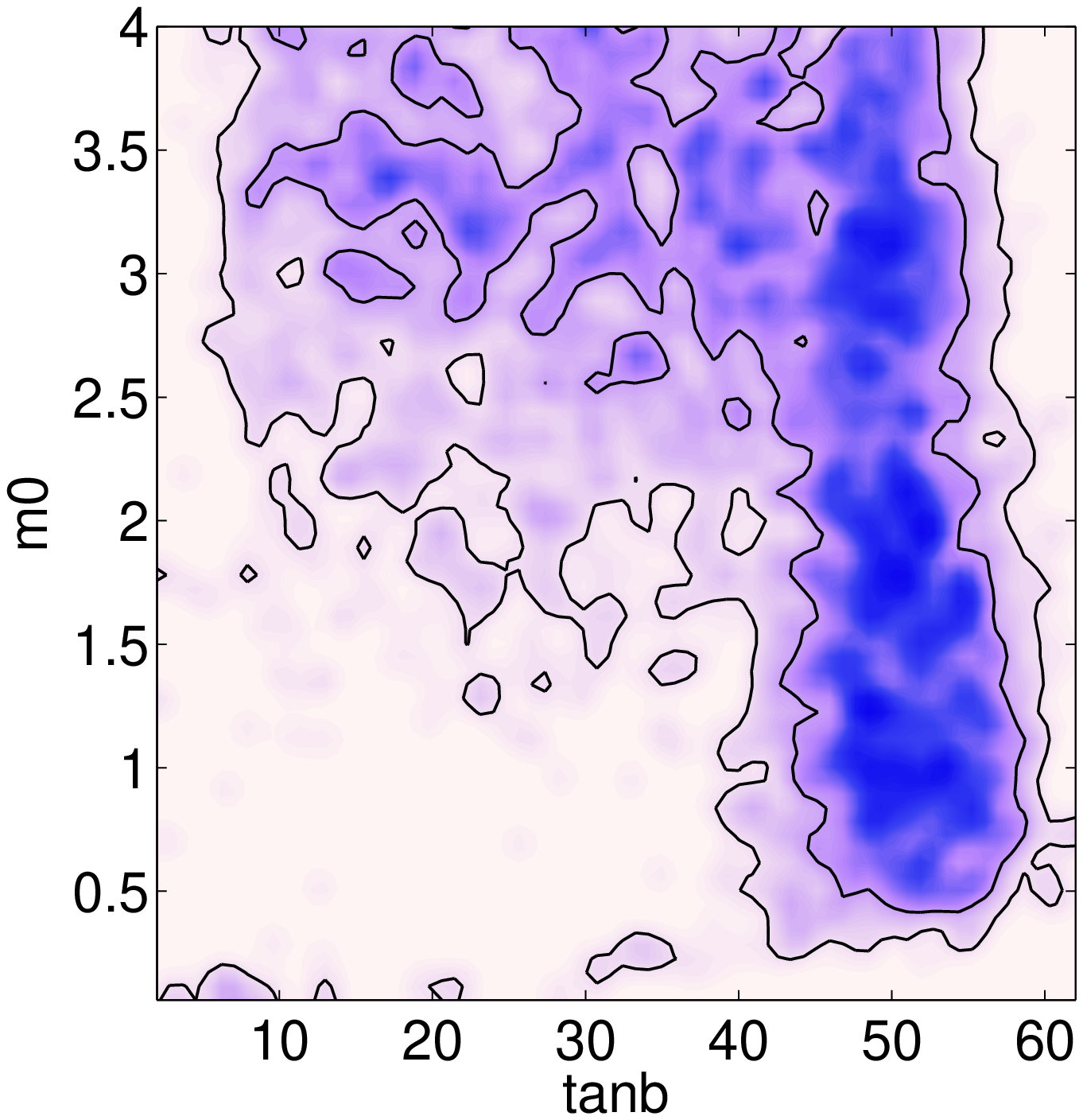}}\\
{\includegraphics[width=0.29\columnwidth,height=0.29\columnwidth]{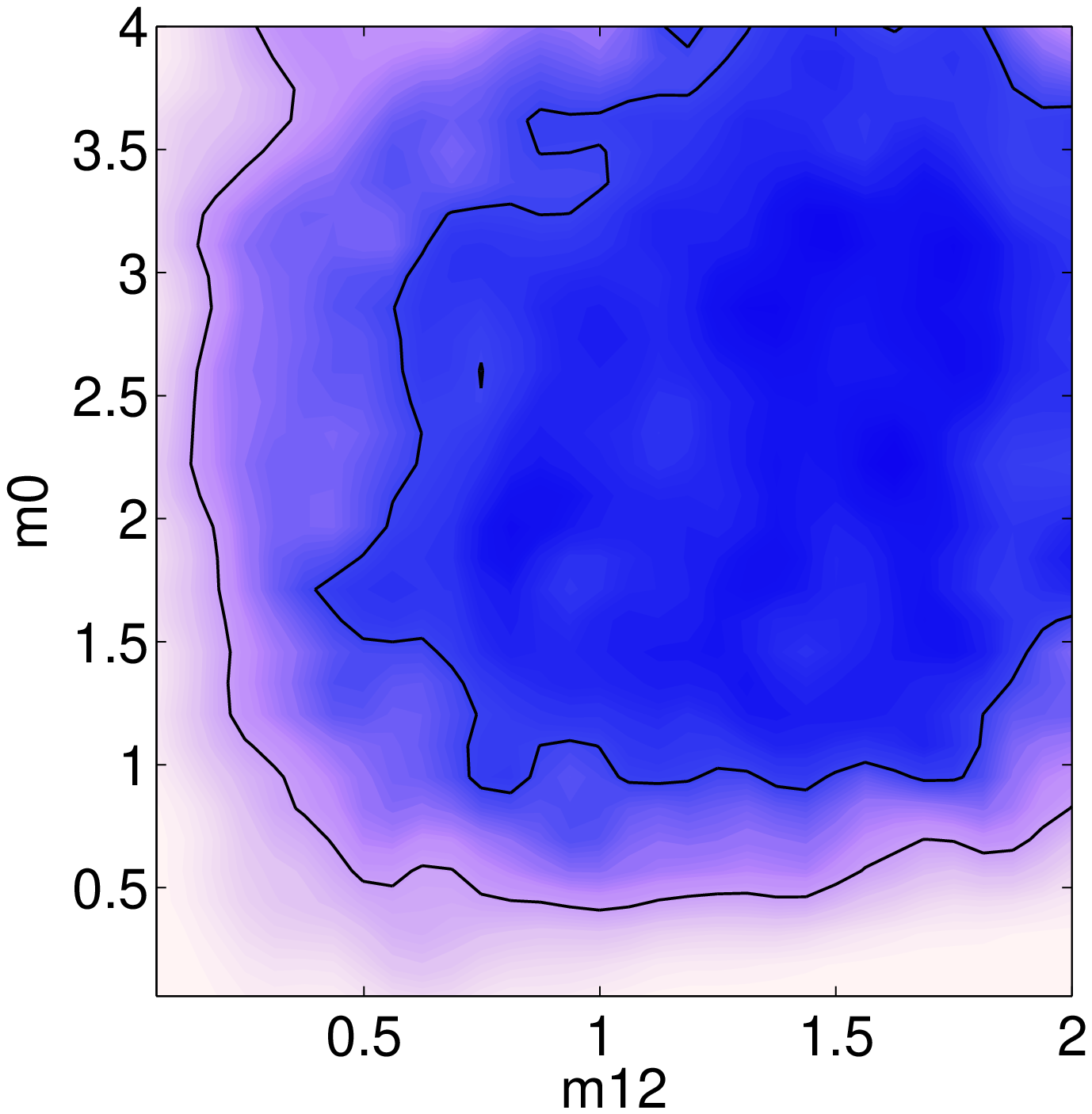}}\hspace{0.3cm}
{\includegraphics[width=0.29\columnwidth,height=0.29\columnwidth]{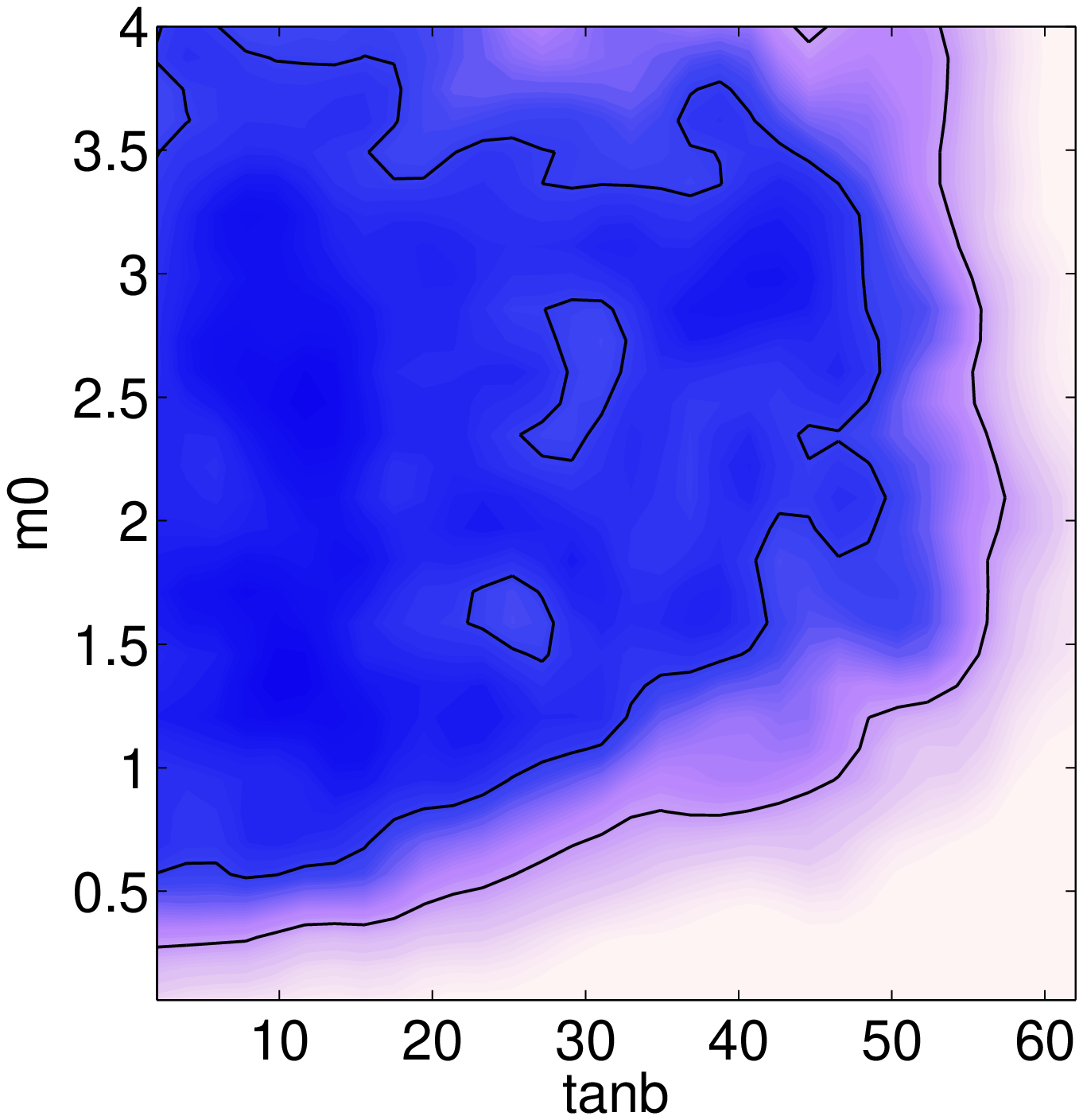}}\\
{\includegraphics[width=0.29\columnwidth,height=0.29\columnwidth]{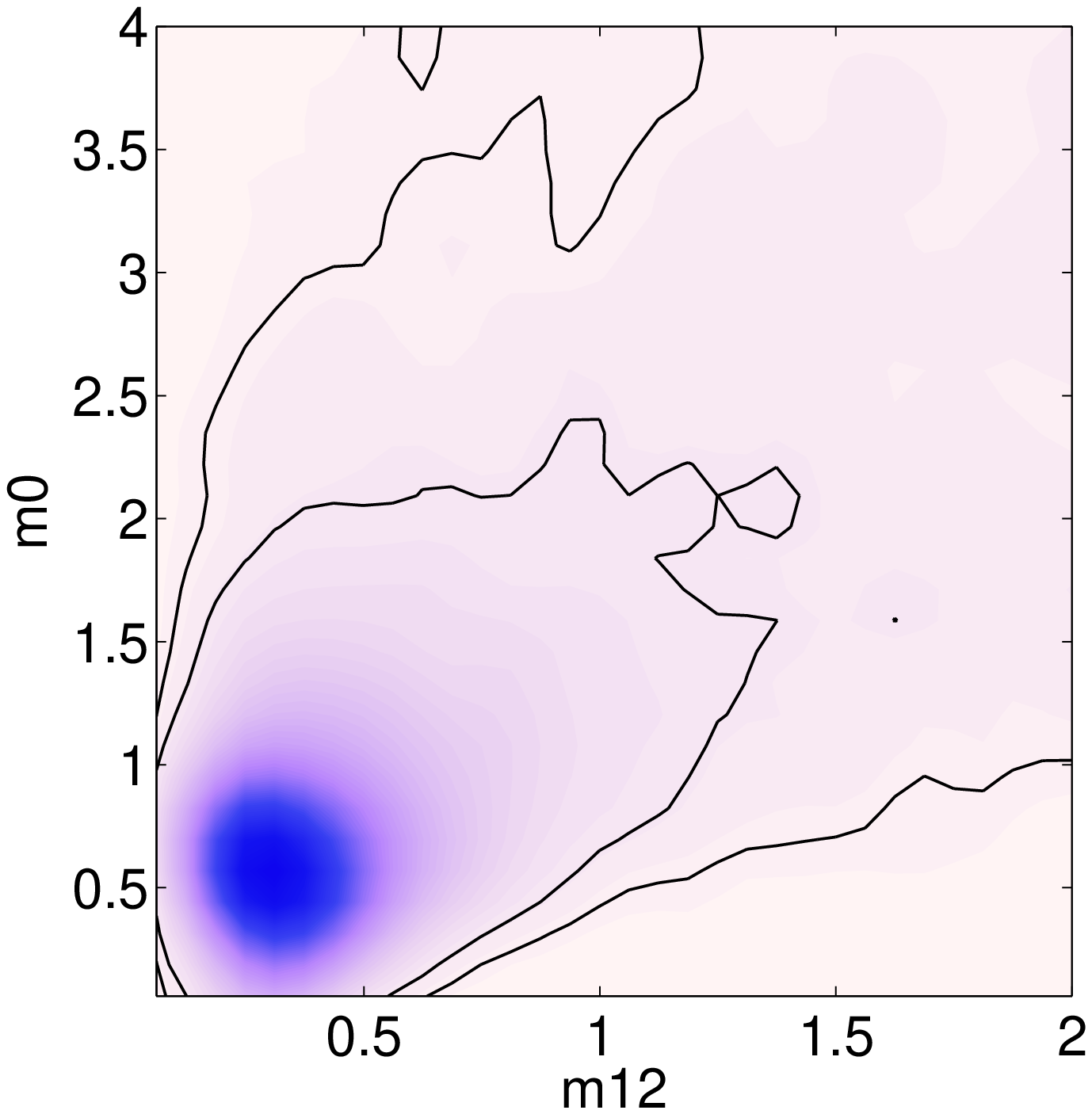}}\hspace{0.3cm}
{\includegraphics[width=0.29\columnwidth,height=0.29\columnwidth]{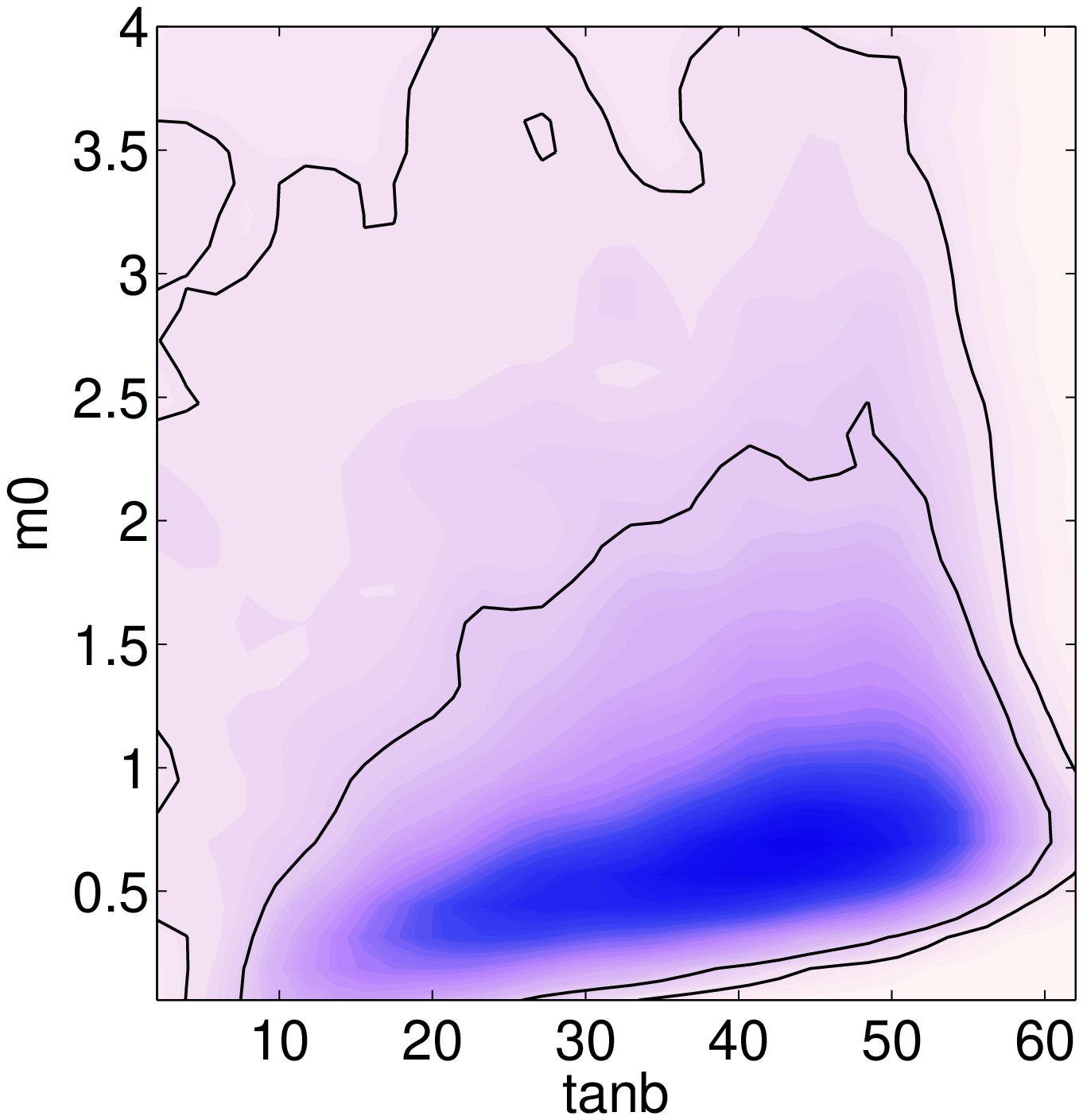}}\\
{\includegraphics[width=0.29\columnwidth,height=0.29\columnwidth]{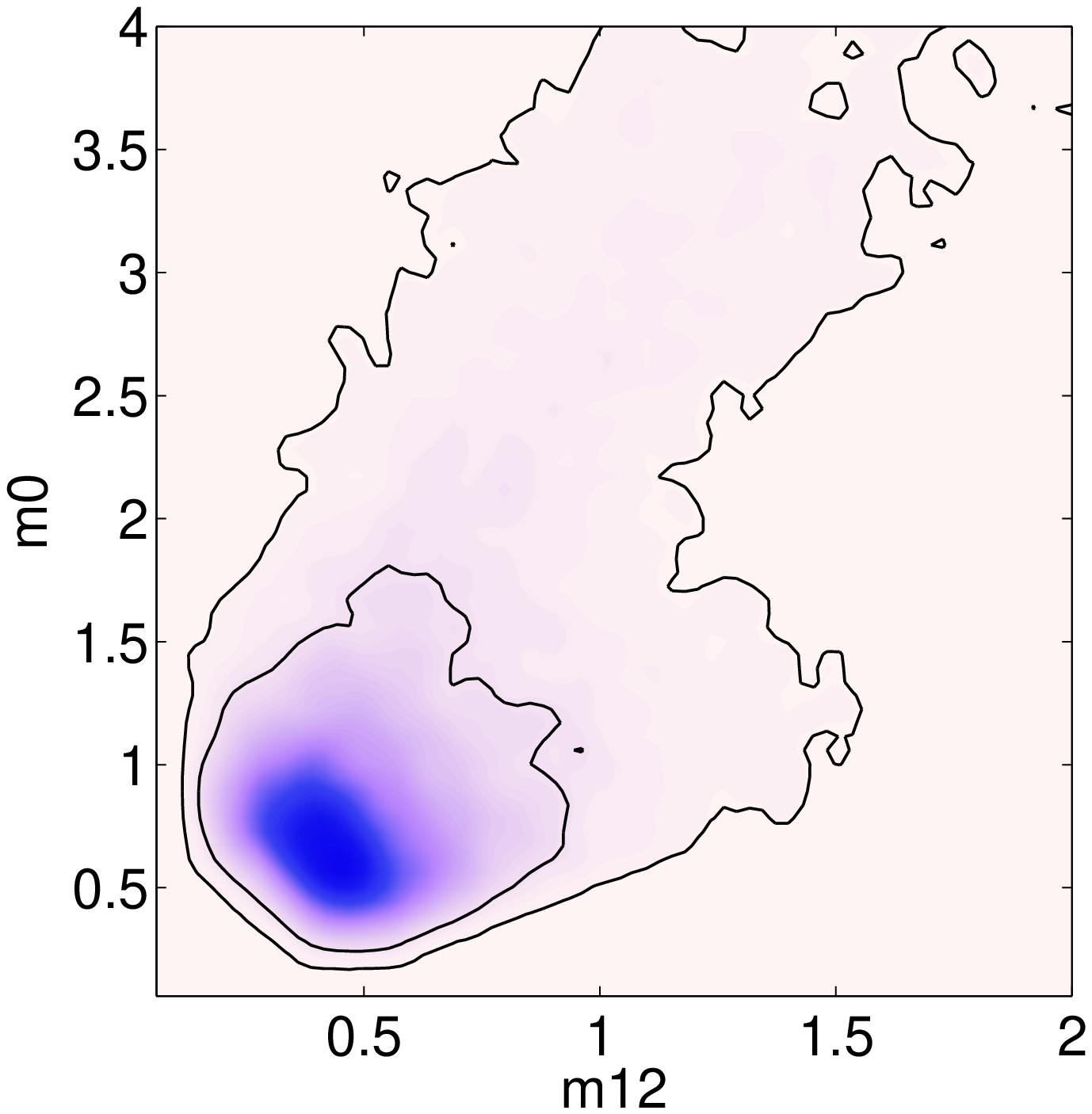}}\hspace{0.3cm}
{\includegraphics[width=0.29\columnwidth,height=0.29\columnwidth]{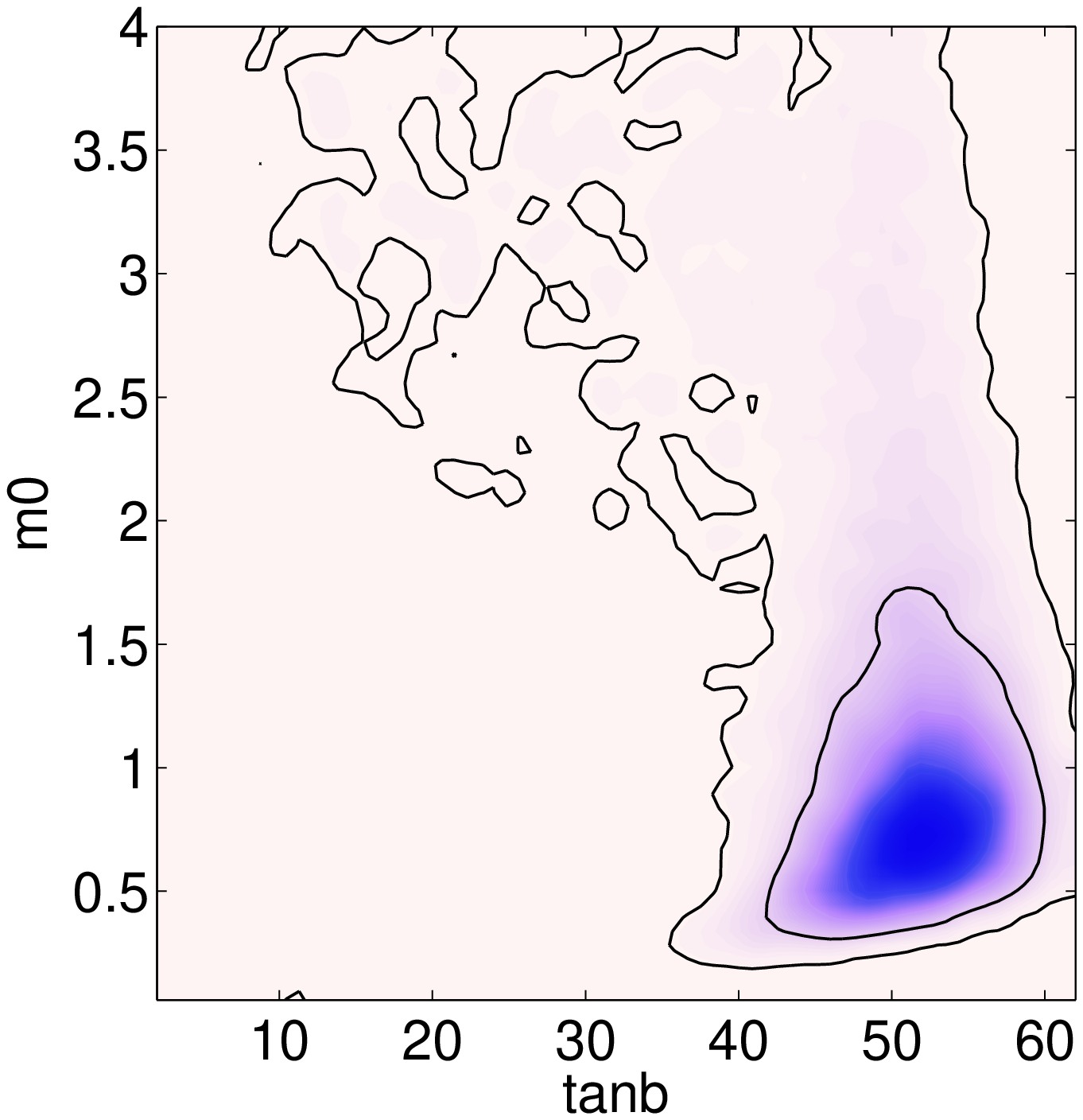}}\\
{\includegraphics[width=0.6\columnwidth, height=0.05\columnwidth]{FIGS/colormap.eps}}
\caption{The 2-dimensional posterior probability distributions of $\mu > 0$ branch of mSUGRA with: from top to
bottom, $\Omega_{\protect\mathrm{DM}}h^2$, $BR(b \rightarrow s \gamma)$, $\delta a_\mu$, and joint analysis of
all three. The inner and outer contours enclose $68$\% and $95$\% of the total probability respectively. All of
the other input parameters in each plane have been marginalised over.}
\label{fig:msugra_consistency_own}}

We now treat all the observables $\tilde{\mathbf{D}}$, apart from $(g-2)_\mu$, $BR(b \rightarrow s \gamma)$ and
$\Omega_{\mathrm{DM}}h^2$ as additional priors on the mSUGRA parameter space in order to see whether these have
any effect on the consistency between $(g-2)_\mu$ and $BR(b \rightarrow s \gamma)$. 
Eq.~\ref{eq:consistency} then becomes: 
\begin{equation}
R=\frac{\Pr((g-2)_\mu, BR(b \rightarrow s \gamma)|\tilde{\mathbf{D}},H_1)}
{\Pr((g-2)_\mu|\tilde{\mathbf{D}},H_0)\Pr(BR(b \rightarrow s \gamma)|\tilde{\mathbf{D}},H_0)},
\label{eq:msugra_consistency_gmu_bsg}
\end{equation}
where the $H_1$ hypothesis states that mSUGRA jointly fits the two observables, whereas $H_0$ states that the
two observables prefer different regions of parameter space.

Since the measurements $D_i$ of the observables used in the likelihood are independent,

\begin{equation}
\Pr((g-2)_\mu, BR(b \rightarrow s \gamma)|\tilde{\mathbf{D}},H_1)=
\frac{\Pr((g-2)_\mu, BR(b \rightarrow s \gamma),\tilde{\mathbf{D}}|H_1)}
{\Pr(\tilde{\mathbf{D}}|H_1)},
\label{eq:msugra_consistency_joint}
\end{equation}
\begin{equation}
\Pr((g-2)_\mu|\tilde{\mathbf{D}},H_0)=
\frac{\Pr((g-2)_\mu,\tilde{\mathbf{D}}|H_0)}
{\Pr(\tilde{\mathbf{D}}|H_0)},
\label{eq:msugra_consistency_gmu}
\end{equation}
\begin{equation}
\Pr(BR(b \rightarrow s \gamma)|\tilde{\mathbf{D}},H_0)=
\frac{\Pr(BR(b \rightarrow s \gamma),\tilde{\mathbf{D}}|H_0)}
{\Pr(\tilde{\mathbf{D}}|H_0)},
\label{eq:msugra_consistency_bsg}
\end{equation}
where $\Pr(\tilde{\mathbf{D}}|H_0)=\Pr(\tilde{\mathbf{D}}|H_1)$ is the Bayesian evidence for the analysis of
$\mu>0$ branch of mSUGRA model with $\tilde{\mathbf{D}}$, all the observables apart from $(g-2)_\mu$, $BR(b
\rightarrow s \gamma)$ and $\Omega_{\mathrm{DM}}h^2$. Hence, to evaluate $R$, we calculate the Bayesian evidence
for the joint as well as individual analysis with $\tilde{\mathbf{D}}$, $(g-2)_\mu$ and $BR(b \rightarrow s
\gamma)$. We evaluate 
\begin{equation}
\log R=0.28 \pm 0.15,
\end{equation}
showing that even the slight inconsistency found between $(g-2)_\mu$, $BR(b \rightarrow s \gamma)$ without
treating $\tilde{\mathbf{D}}$ as additional priors on mSUGRA model, has now vanished which means that
$\tilde{\mathbf{D}}$ data-sets have cut-off the discrepant regions of the two constraints.

\section{Summary and Conclusions}\label{sec:summary}

Bayesian analysis methods have been used successfully in astronomical
applications~\cite{McLachlan,Marshall,Slosar,Mukherjee,Basset,Trotta,Beltran,Bridges,trotta07,trotta08}.
However, the application of Bayesian methods to problems in particle physics is less established, due perhaps to
the highly degenerate and multi-modal parameter spaces which present a great difficulty for the standard MCMC
based techniques. Bank sampling~\cite{bank} provides a practical means of MCMC parameter estimation and evidence
{\em ratio}\/ estimation under such circumstances, but it cannot calculate the evidence itself. We have shown
that the {\sc MultiNest} technique not only handles these complex distributions in a highly efficient manner but
also allows the calculation of the Bayesian evidence enabling one to perform the model comparison. This could be
of great importance in distinguishing different beyond the Standard Model theories, once high quality data from
the LHC becomes available. 

Our central results are summarised in Table~\ref{tab:prob_odds}. It is clear that, in global mSUGRA fits to
indirect data, $\mu > 0$ is somewhat preferred to $\mu<0$, mainly due to data from the anomalous magnetic moment
of the muon, which outweighs the preference for $\mu < 0$ from the measured branching ratio of a $b$ quark into
an $s$ quark and a photon and the SM prediction when some of the NNLO QCD contributions are included. For a given
measure and range of the prior, the evidence ratio between the different signs of $\mu$ is accurately determined
by the {\sc MultiNest} technique. Despite additional data from the $b-$sector and the anomalous magnetic moment
of the muon having a higher discrepancy with the Standard Model prediction, there is still not enough power in
the data to make the fits robust enough. We see a signal of this in the fact that the evidence ratio $P_+/P_-$ is
highly dependent upon the measure and range of the prior distribution of mSUGRA parameters. We obtain
$P_+/P_-=6-61$ depending upon which range and which measure is chosen. All of these values exhibit positive
evidence, but on the scale summarised in Table~\ref{tab:Jeffreys}, `weak' evidence is characterised as being bigger
than $3$, `moderate' as bigger than $12$. Thus we cannot unambiguously conclude that the evidence is strongly in
favour of $\mu>0$: only weak. A further test also suggested that within one prior measure and range, and for
$\mu>0$, the tension between the observables $(g-2)_\mu$ and $BR(b\rightarrow s \gamma)$ is not statistically
significant.

\appendix

\section{Consistency Check with Bayesian Evidence}\label{app:cosistency}
In order to motivate the use of Bayesian evidence to quantify the consistency between different data-sets as
discussed in Section~\ref{sec:bayesian}, we apply the method to the classic problem of fitting a straight line
through a set of data points.

\subsection{Toy Problem}
\label{app:cosistency:toy}
We consider that the true underlying model for some process is a straight line described by:
\begin{equation}
y(x)=mx+c,
\label{eq:line}
\end{equation}
where $m$ is the slope and $c$ is the intercept. We take two independent sets of measurements $\mathbf{D}_1$ and
$\mathbf{D}_2$ each containing $5$ data points. The $x$ value for all these measurements are drawn from a uniform
distribution $\mathcal{U}(0,1)$ and are assumed to be known exactly.

\subsubsection{Case I: Consistent Data-Sets}
\label{app:cosistency:toy:consistent}

\FIGURE{
\includegraphics[width=0.4\columnwidth,height=0.4\columnwidth]{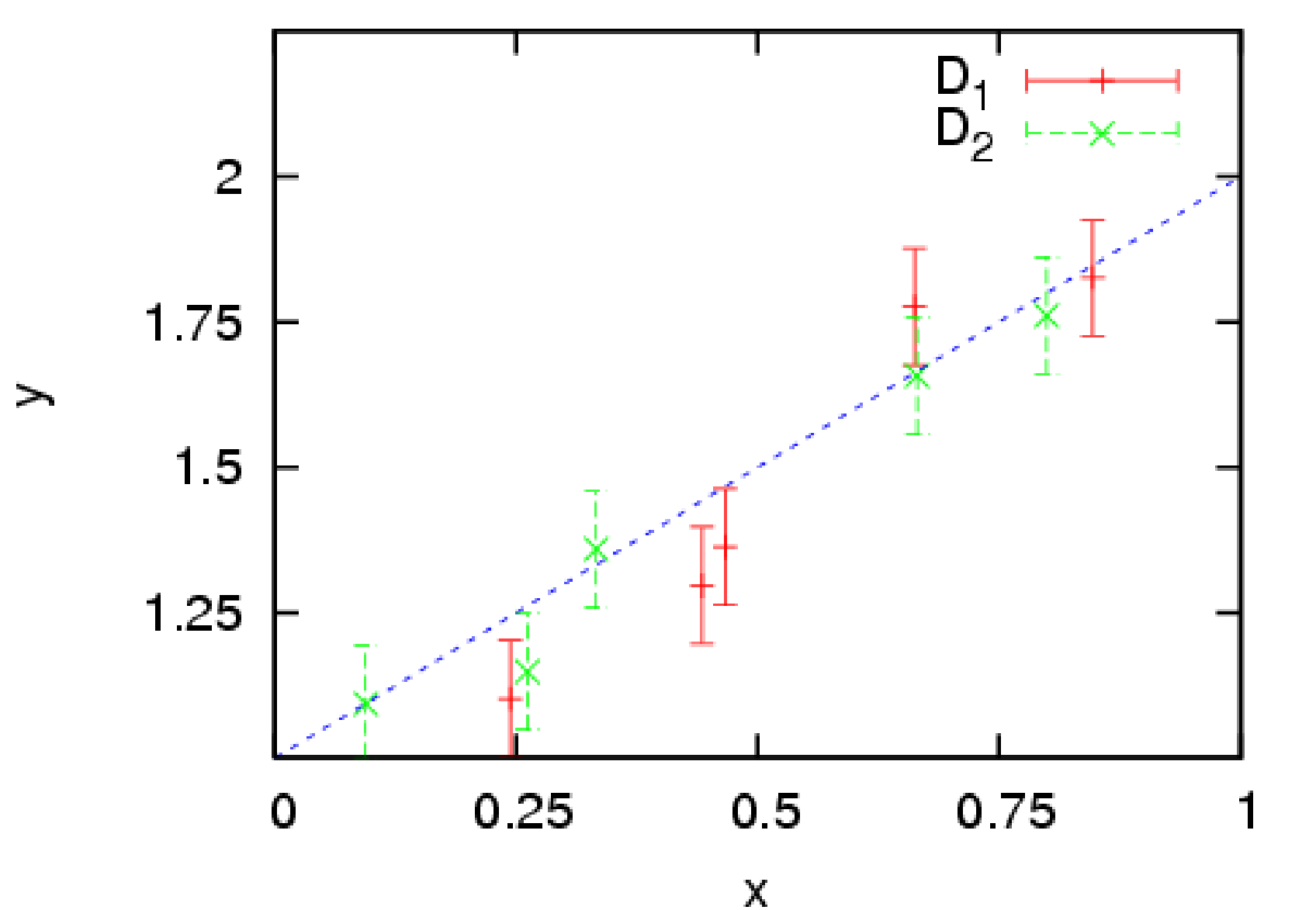}
\includegraphics[width=0.38\columnwidth,height=0.38\columnwidth,angle=0]{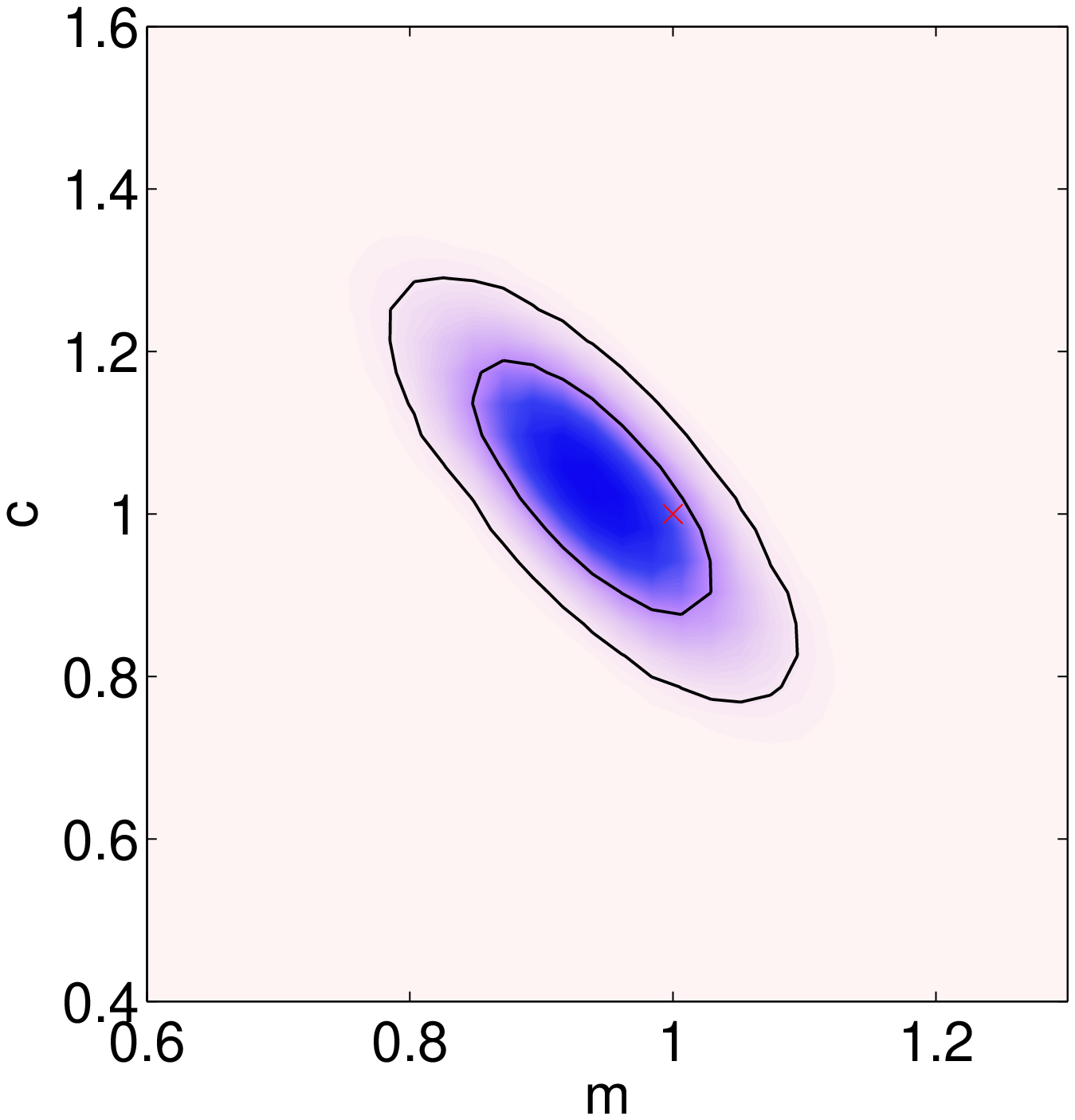}
\includegraphics[width=0.38\columnwidth,height=0.38\columnwidth,angle=0]{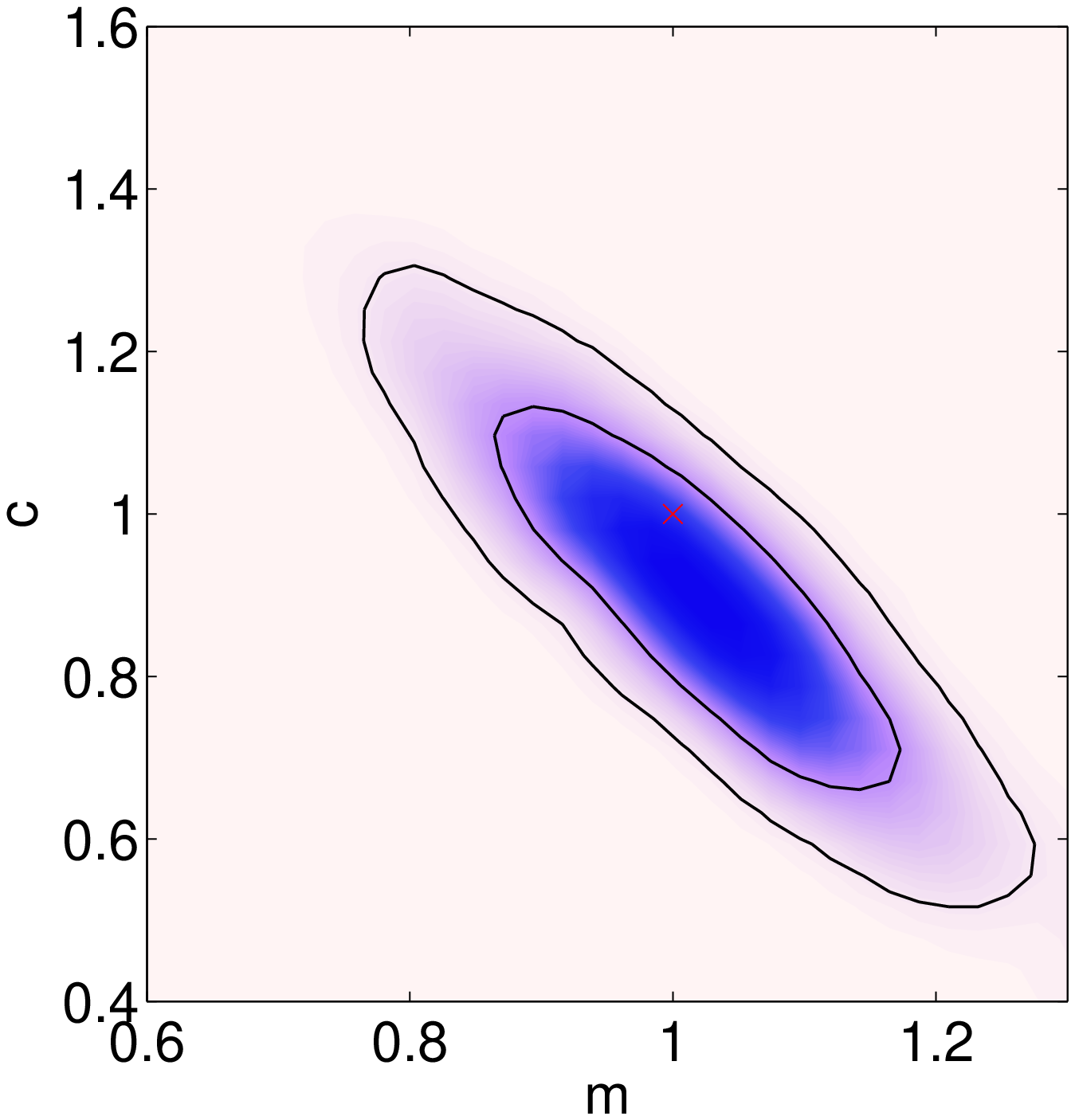}
\includegraphics[width=0.38\columnwidth,height=0.38\columnwidth,angle=0]{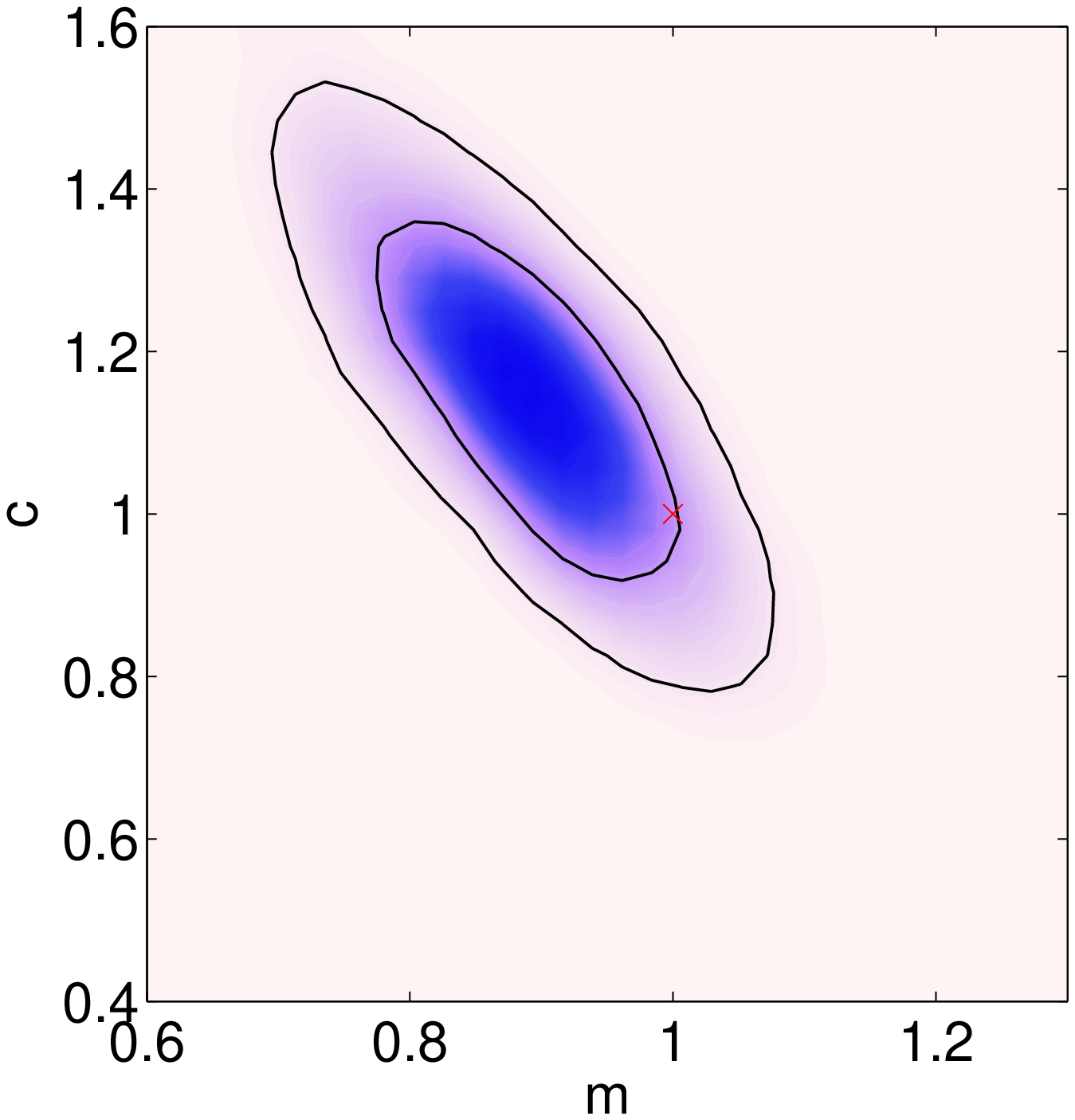}
\caption{Upper left: Data-sets $\mathbf{D}_1$ and $\mathbf{D}_2$ drawn from a straight line model (solid line)
with slope $m=1$ and intercept $c=1$ and subject to independent Gaussian noise with root mean square $\sigma_1 =
\sigma_2 = 0.1$. Upper right: Posterior $\Pr(m,c|\mathbf{D},H_1)$ assuming that data-sets $\mathbf{D}_1$ and
$\mathbf{D}_2$ are consistent. Lower left: Posterior $\Pr(m,c|\mathbf{D},H_1)$ for data-set $\mathbf{D}_1$.
Lower right: Posterior $\Pr(m,c|\mathbf{D},H_1)$ for data-set $\mathbf{D}_2$. The inner and outer contours
enclose $68$\% and $95$\% of the total probability respectively. The true parameter value is indicated by red
crosses.}
\label{fig:app:line1}}

In the first case we consider $m=1$, $c=1$ and add Gaussian noise with standard deviation $\sigma_1 = 0.1$ and
$\sigma_2 = 0.1$ for data-sets $\mathbf{D}_1$ and $\mathbf{D}_2$ respectively. Hence both the data-sets provide
consistent information on the underlying process.

We assume that the errors $\sigma_1$ and $\sigma_2$ on the data-sets $\mathbf{D}_1$ and $\mathbf{D}_2$ are known
exactly. The likelihood function can then be written as:

\begin{equation}
\mathcal{L}(m,c) \equiv \Pr(\mathbf{D}|m,c,H) = \prod_{i}^{} \Pr(\mathbf{D_i}|m,c, H),
\label{eq:app:like1}
\end{equation}
where
\begin{equation}
\Pr(D_i|m,c, H)=\frac{1}{\sqrt{2\pi \sigma_i^2}}\exp[-\chi_i^2/2]
\label{eq:app:like2}
\end{equation}
and
\begin{equation}
\chi_i^2=\sum_{j}^{} \frac{(y(x_j)-\tilde{y}(x_j))^2}{\sigma_i^2}.
\label{eq:like3}
\end{equation}
where $\tilde{y}(x_j)$ is the predicted value of $y$ at a given $x_j$.

We impose uniform, $\mathcal{U}(0,2)$ priors on both $m$ and $c$. In Fig.~\ref{fig:app:line1} we show the data points and
the posterior for the analysis assuming the data-sets $\mathbf{D}_1$ and $\mathbf{D}_2$ are consistent. The true
parameter value clearly lies inside the contour enclosing $68$\% of the posterior probability.

In order to quantify the consistency between the data-sets $\mathbf{D}_1$ and $\mathbf{D}_2$, we evaluate $R$ as
given in Eq.~\ref{eq:consistency} which for this case becomes: 
\begin{equation}
R=\frac{\Pr(\mathbf{D}_1,\mathbf{D}_2|H_1)}{\Pr(\mathbf{D}_1|H_0)\Pr(\mathbf{D}_2|H_0)},
\label{eq:app:consistency}
\end{equation}
where the $H_1$ hypothesis states that the model jointly fits the data-sets $\mathbf{D}_1$ and $\mathbf{D}_2$,
whereas $H_0$ states that $\mathbf{D}_1$ and $\mathbf{D}_2$ prefer different regions of parameter space. We
evaluate, 
\begin{equation}
\log R=3.2 \pm 0.1,
\end{equation}
showing strong evidence in favour of $H_1$. 

\subsubsection{Case II: Inconsistent Data-Sets}
\label{app:cosistency:toy:inconsistent}

\FIGURE{
\includegraphics[width=0.38\columnwidth,height=0.38\columnwidth]{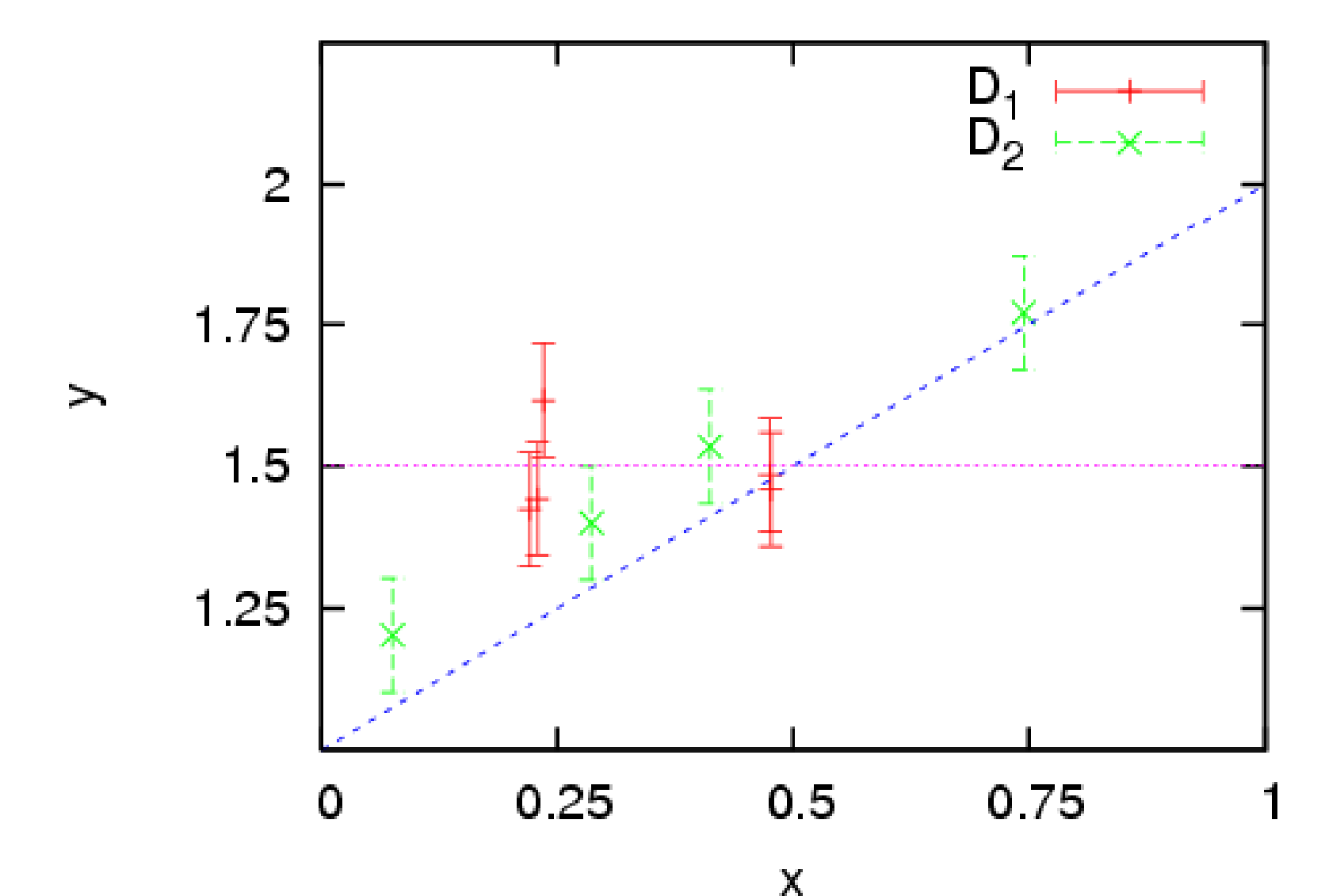}
\includegraphics[width=0.38\columnwidth,height=0.38\columnwidth]{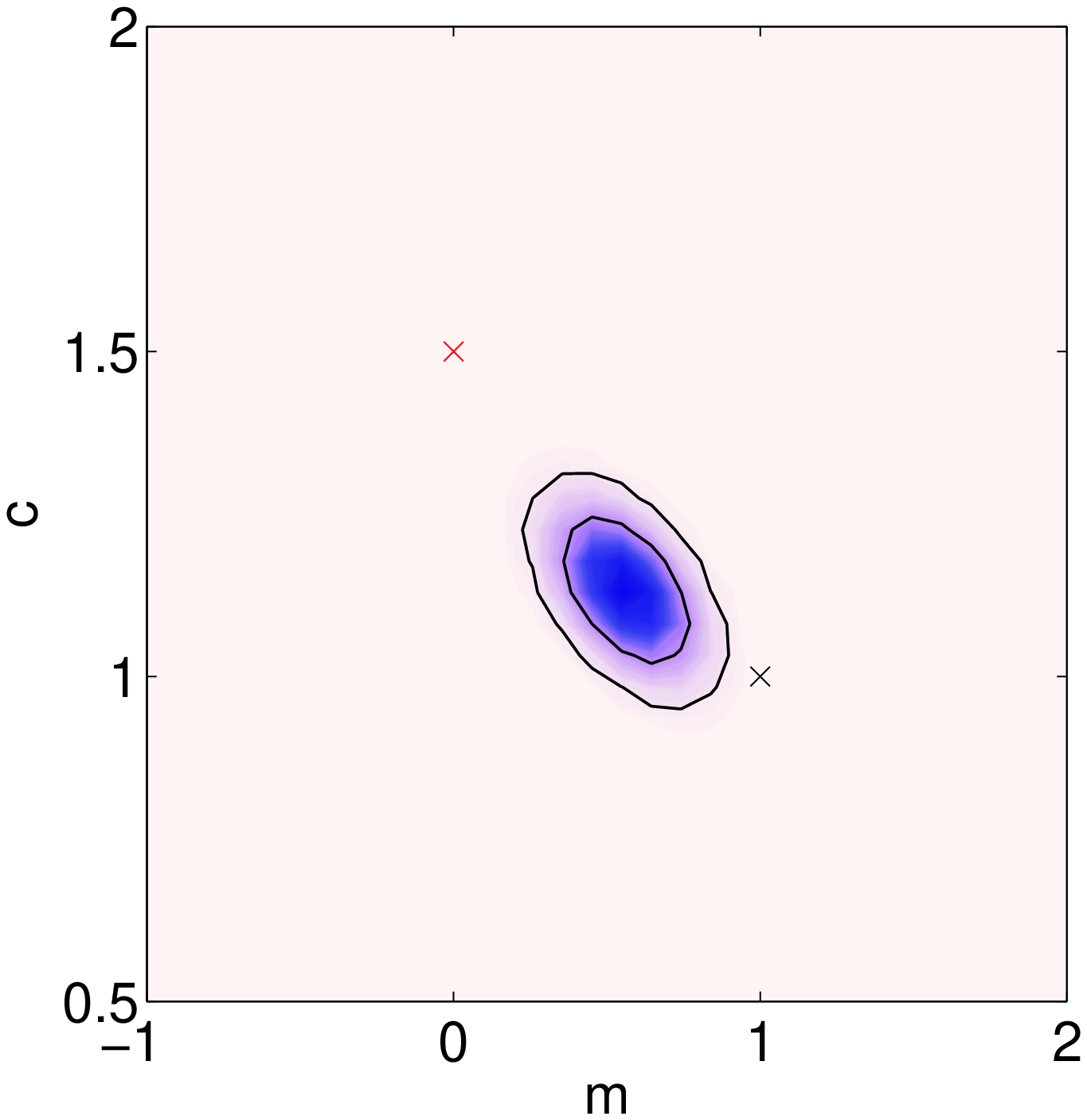}
\includegraphics[width=0.38\columnwidth,height=0.38\columnwidth,angle=0]{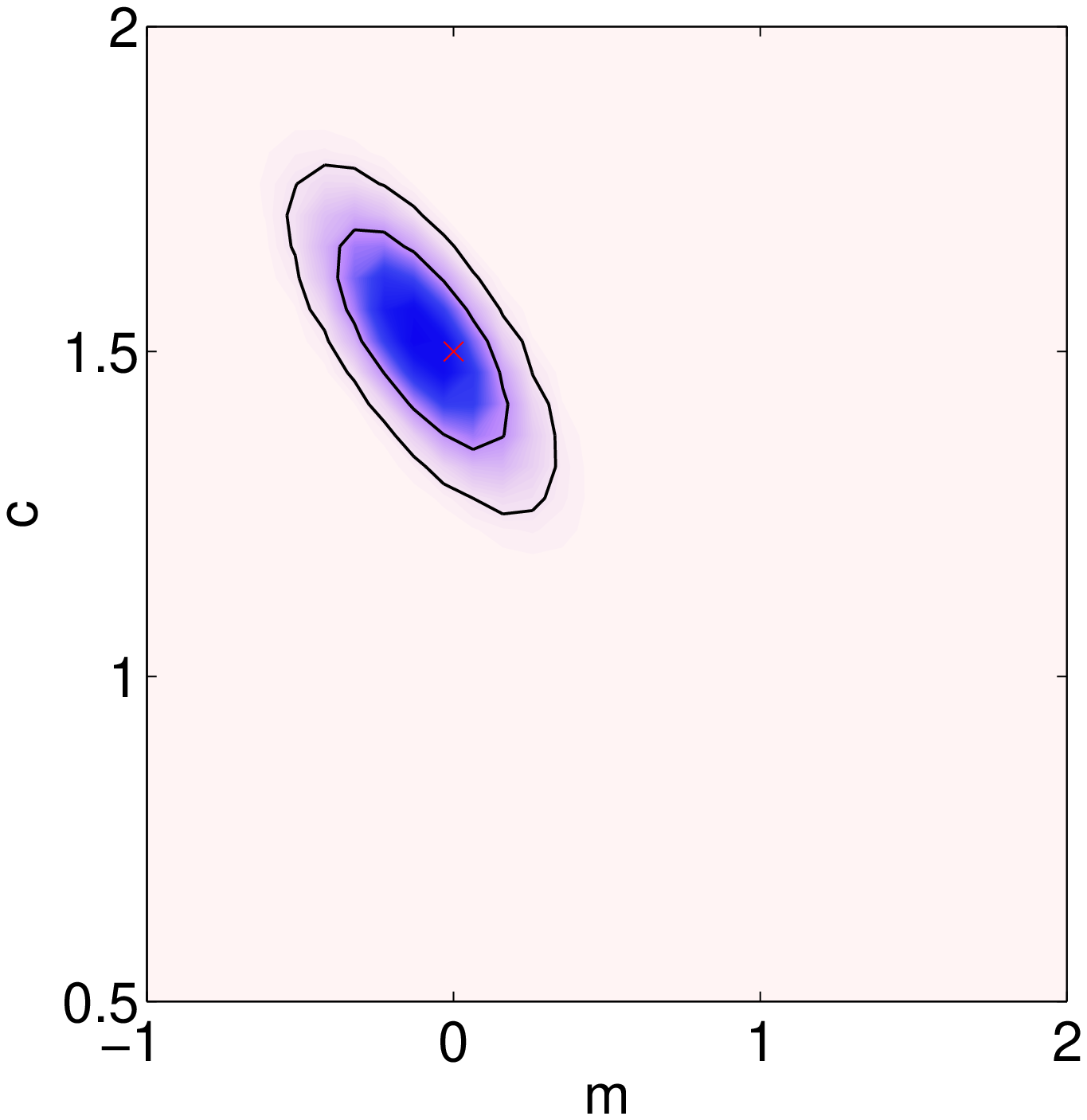}
\includegraphics[width=0.38\columnwidth,height=0.38\columnwidth,angle=0]{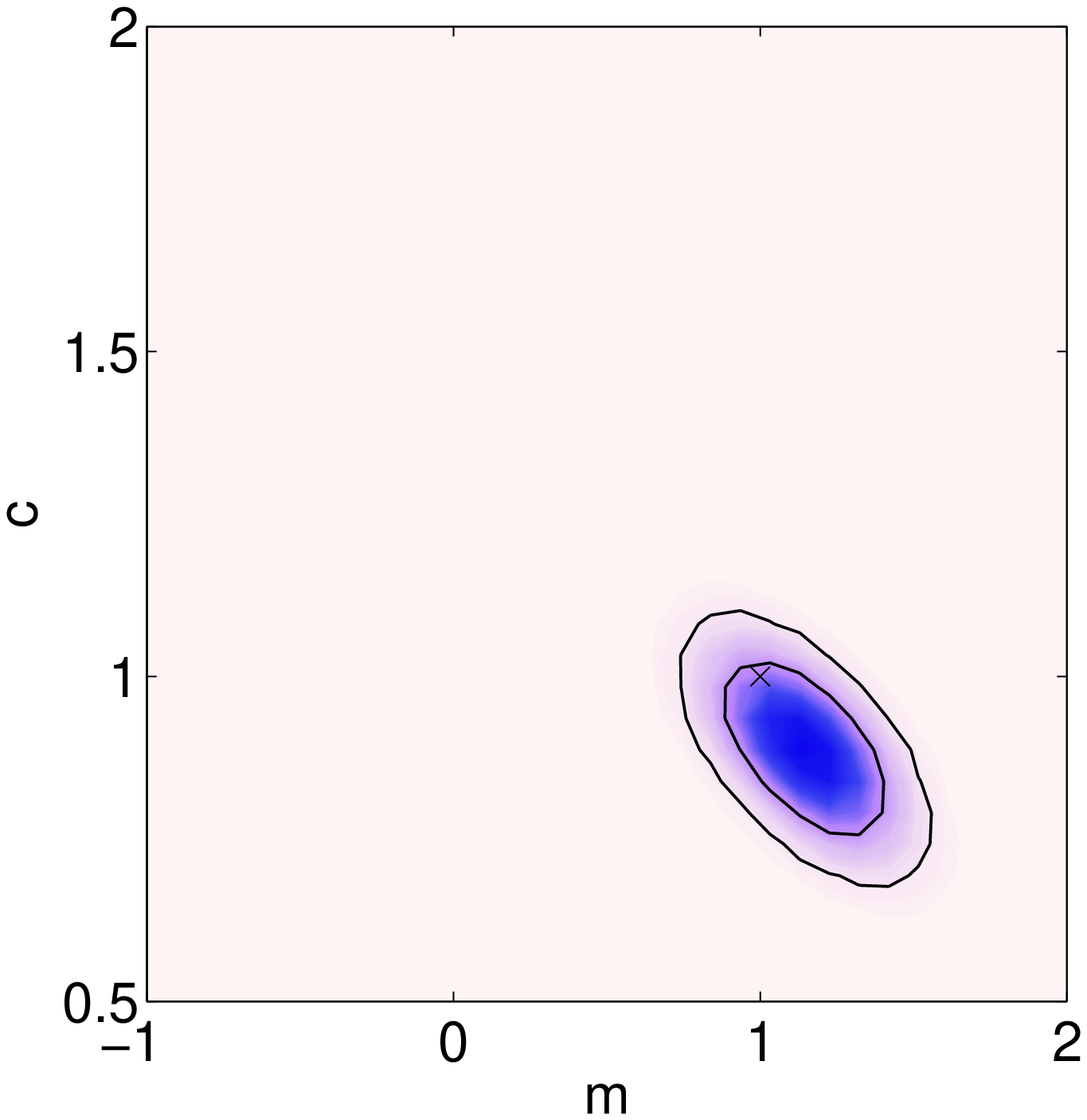}
\caption{Upper left: Data-sets $\mathbf{D}_1$ and $\mathbf{D}_2$ drawn from a straight line model (solid line)
with slope $m=0$, $c=1.5$ and $m=1$, $c=1$ respectively and subject to independent Gaussian noise with root mean
square $\sigma_1 = \sigma_2 = 0.1$. Upper right: Posterior $\Pr(m,c|\mathbf{D},H_1)$ assuming that data-sets
$\mathbf{D}_1$ and $\mathbf{D}_2$ are consistent. Lower left: Posterior $\Pr(m,c|\mathbf{D},H_1)$ for data-set
$\mathbf{D}_1$. Lower right: Posterior $\Pr(m,c|\mathbf{D},H_1)$ for data-set $\mathbf{D}_2$. The inner and
outer contours enclose $68$\% and $95$\% of the total probability respectively. The true parameter values are
indicated by red and black crosses for Data-sets $\mathbf{D}_1$ and $\mathbf{D}_2$ respectively.}
\label{fig:app:line2}}

We now introduce systematic error into the data-set $\mathbf{D}_1$ by drawing from an incorrect straight line
model with $m=0$ and $c=1.5$. Measurements for $\mathbf{D}_2$ are still drawn from a straight line with $m=1$ and
$c=1$. We assume that the errors $\sigma_1=0.1$ and $\sigma_2=0.1$, for $\mathbf{D}_1$ and $\mathbf{D}_2$
respectively, are both quoted correctly. 

We impose uniform priors, $\mathcal{U}(-1,2)$ and $\mathcal{U}(0,2)$, on $m$ and $c$ respectively. In Fig.~\ref{fig:app:line2} we
show the data points and the posterior for the analysis assuming the data-sets $\mathbf{D}_1$ and $\mathbf{D}_2$
are consistent as well as for the analysis with data-sets $\mathbf{D}_1$ and $\mathbf{D}_2$ taken separately.
In spite of the fact that the two sets of true parameter values define a direction along the natural degeneracy
line in the $(m,c)$ plane, neither of the true parameter values lie inside the contour enclosing $95$\% of the
posterior probability. Also, it can be seen that the there is no overlap between the posteriors for data-sets
$\mathbf{D}_1$ and $\mathbf{D}_2$ and so both models can be excluded at a high significance level. We again
compute $R$ as given in Eq.~\ref{eq:app:consistency} and evaluate it to be, 
\begin{equation}
\log R=-13.1 \pm 0.1,
\end{equation}
showing evidence in favour of $H_0$ i.e.\ the data-sets $\mathbf{D}_1$ and $\mathbf{D}_2$ provide inconsistent
information on the underlying model.

\appendix

\section*{Acknowledgements}
We thank Nazila Mahmoudi for her help with {\tt SuperIso2.0} and Pietro Slavich for advice about the $b
\rightarrow s \gamma$ calculation. This work has been partially supported by STFC\@ and the EU FP6 Marie Curie
Research \& Training Network ``UniverseNet'' (MRTN-CT-2006-035863). The computation was carried out largely on
the {\sc Cosmos} UK National Cosmology Supercomputer at DAMTP, Cambridge and the Cambridge High Performance
Computing Cluster Darwin and we thank Victor Travieso, Andrey Kaliazin and Stuart Rankin for their computational
assistance. FF is supported by the Cambridge Commonwealth Trust, the Isaac Newton Trust and the Pakistan Higher
Education Commission Fellowships. SSA is supported by the Gates Cambridge Trust. RT is supported by the Lockyer
Fellowship of the Royal Astronomical Society, St Anne's College, Oxford and STFC. The authors would like to thank
the European Network of Theoretical Astroparticle Physics ENTApP ILIAS/N6 under contract number
RII3-CT-2004-506222 for financial support.

\label{lastpage}

\end{document}